\documentstyle[12pt,epsf]{article}

\advance \topmargin by -\headheight
\advance \topmargin by -\headsep

\evensidemargin \oddsidemargin
 \def\ex{{\hbox{\rm e}}}

  \def\tr{{\hbox{\rm Tr}}}

 \def\vev{vacuum expectation value}

\def\ie{{\em i.e.}}

\def\np{Nucl. Phys.}

\def\tam{Trans. Am. Math. Soc.}

\def\bams{Bull. AMS}
\def\am{Ann. of Math.}

\def\jpsc{J. Phys. Soc. Jap.}
\def\topo{Topology}

\def\knot{Journal of Knot Theory and Its Ramifications}

\newcommand{\RR}{{\mbox{{\bf R}}}}
\newcommand{\RRs}{{\mbox{{\small \bf R}}}}

\newcommand{\beq}{\begin{equation}}
\newcommand{\eeq}{\end{equation}}
\newcommand{\bear}{\begin{eqnarray}}
\newcommand{\eear}{\end{eqnarray}}
\newcommand{\W}{{\cal W}}
\newcommand{\F}{{\cal F}}
\newcommand{\x}{{\cal O}}
\newcommand{\LL}{{\cal L}}

\def\calw{{\cal W}}
\def\calz{{\cal Z}}

\def\calc{{\cal C}}

\def\dr{\dot R_}

\def\ds{\dot s_}
\def\da{\dot A_}
\def\dga{\dot \gamma_}
\def\ga{\gamma_}

\def\cls{{closing}}
\def\vev{vacuum expectation value}
\def\tr{{\rm Tr}}

\def\too{\longrightarrow}

\newfont{\namefont}{cmr10}
\newfont{\addfont}{cmti7 scaled 1440}
\newfont{\headfontb}{cmbx10 scaled 1728}
\renewcommand{\theequation}{{\rm \thesection.\arabic{equation}}}
\begin{document}
\begin{titlepage}
\begin{center} {\headfontb Vassiliev Invariants for Links\\ ~from
Chern-Simons Perturbation Theory}
\footnote{This work is supported in part by funds provided by the
U.S.A. DOE under cooperative research agreement
\#DE-FC02-94ER40818 and by the DGICYT of Spain under grant PB93-0344.}
\end{center}
\vskip 0.3truein
\begin{center} {\namefont M. Alvarez\footnote{Email: marcos@mitlns.mit.edu}}
\end{center}
\begin{center} {\addfont{Center for Theoretical Physics,}}\\
{\addfont{Massachusetts Institute of Technology}}\\ {\addfont{Cambridge,
Massachusetts 02139 U.S.A.}}
\end{center}
\vskip 0.3truein
\begin{center} {\namefont J.M.F. Labastida\footnote{Email:
labastida@gaes.usc.es} and E. P\'erez}
\end{center}
\begin{center} {\addfont{Departamento de F\'\i sica de Part\'\i culas,}}\\
{\addfont{Universidade de Santiago}}\\ {\addfont{E-15706 Santiago de
Compostela, Spain}}
\end{center}
\vskip 1truein

\begin{center}
\bf ABSTRACT
\end{center} 

The general structure of the perturbative expansion of the vacuum expectation
value of a product of Wilson-loop operators is analyzed in the context of 
Chern-Simons gauge theory. Wilson loops are opened into Wilson lines in order to
unravel the algebraic structure encoded in the group factors of the perturbative
series expansion. In the process a factorization theorem is proved for Wilson
lines. Wilson lines are then closed back into Wilson loops and new link
invariants of finite type are defined. Integral expressions for these invariants
are presented for the first three primitive ones of lower degree in the case of
two-component links. In addition, explicit numerical results are obtained for
all two-component links of no more than six crossings up to degree four.

\vskip3.5truecm
\leftline{MIT-CTP-2547  \hfill June 1996}
\leftline{USC-FT-30-96}
\leftline{hep-th/9607030}
\smallskip
\end{titlepage}
\setcounter{footnote}{0}

\setcounter{equation}{0}
\section{Introduction}

The complete classification of knots and links embedded in three dimensional 
manifolds is still an open problem. Apart from the classical results of 
Alexander, Reidemeister and others (see, for example, 
\cite{reidemeister,rolfsen}), we have now the polynomial invariants of Jones
\cite{jones} and its generalizations \cite{homfly,kauf,aw}. Unlike the classical 
Alexander polynomial, these polynomials are able to distinguish knots or links 
from their mirror images. However, it is still not known if they separate knots. 

All these new invariants are strongly rooted in ideas and methods of Quantum
Field Theory or Statistical Mechanics. By using Yang-Baxter models, it is 
possible to define the Jones polynomial and its relatives, which can also 
be described in terms of quantum groups \cite{restu}. On the other hand, the
formalism of Chern-Simons gauge theory leads in a natural way to all these new 
invariants, each of them corresponding to a choice of the gauge group 
\cite{witten}. The main observable in Chern-Simons theory is the Wilson-loop
operator, which for a given gauge group depends only on the knot class of 
the loop. A non-perturbative evaluation of the \vev\ of this operator leads
directly to the above mentioned polynomial invariants. 

A different set of invariants are the Vassiliev invariants. These were 
first proposed  in \cite{vassi,vassidos} to classify knot types. To each knot 
corresponds an infinite sequence of rational numbers which have to satisfy 
some consistency conditions in order to be knot class invariants. This 
infinite sequence is divided into finite subsequences, which form vector 
spaces. Each subsequence is indexed by a positive integer called its 
order. The number of independent elements in each finite subsequence is called 
the dimension of the space of Vassiliev invariants at that order. 

An axiomatic definition of these invariants was formulated in 
\cite{birlin,birman}, in terms of inductive relations for singular knots. This
approach is best suited to show the relation to other knot invariants based on 
quantum groups or in Chern-Simons gauge theory 
\cite{drorcon,birlin,birman,drortesis,lin}. Several works have been
performed to analyze Vassiliev invariants in both frameworks 
\cite{konse,drortopo,piuni,numbers,haya}. In \cite{konse,drortopo} it was shown
that Vassiliev invariants can be understood in terms of representations of
chord diagrams whithout isolated chords modulo the so called 4T relations
(weight systems), and that using semi-simple Lie algebras weight systems can
be constructed. It was also shown in \cite{drortopo}, using Kontsevitch's
representation for Vassiliev invariants \cite{konsedos}, that the space of
weight systems is the same as the space of Vassiliev invariants. In 
\cite{piuni} it was argued that these representations are precisely the ones 
underlying quantum-group or Chern-Simons invariants. 

The connection of Vassiliev invariants to Chern-Simons theory
shows up through a perturbative evaluation of the \vev\ of the Wilson-loop 
operators, in the sense of ordinary perturbative Quantum Field Theory. We 
observed in \cite{numbers} that the generalization of 
the integral or geometrical knot invariant first proposed in \cite{gmm} and 
further analyzed in \cite{drortesis}, as well as the invariant itself, are 
Vassiliev invariants. These invariants arise naturally in the perturbative 
analysis of the Wilson loop. In \cite{numbers} we proposed an organization of 
those geometrical invariants and we described a procedure for their calculation
from known polynomial knot invariants. This procedure has been applied to
obtain Vassiliev knot invariants up to order six for all prime knots up
to six crossings \cite{numbers} and for all torus knots \cite{torus}. 
These geometrical invariants have also been studied by
Bott and Taubes \cite{bot} using a different approach. The relation of this
approach to the one in \cite{numbers} has been studied recently 
in \cite{altfr}.

The Vassiliev invariants of a given knot 
form an algebra in the sense that the product of two invariants of orders $i$
and $j$ is an invariant of order $i+j$. Therefore the set of independent 
Vassiliev invariants at a given order can be divided into two subsets: those 
that are products of invariants of lower orders (composite invariants), and
those that are not (primitive invariants). In \cite{factor} we called this
property ``factorization'', and showed how it can be exploited to express the
\vev\ of the Wilson-loop operator associated with the knot as an exponential 
whose argument includes only primitive invariants. This was accomplished by
choosing a particular kind of basis of group factors that we called 
``canonical''. 

The aim of this paper is to extend the formalism in \cite{numbers,factor} to 
two-component links. A straightforward application of the formalism of 
canonical bases, though feasible, would not be satisfactory due to the fact 
that in the case of links there is not a simple algebraic structure among 
group factors similar to the one present in the case of knots. A similar 
algebraic structure appears, however, when open links are considered.  In the 
resulting framework a factorization theorem as the one presented for knots in
\cite{factor} holds. With the help of this theorem finite type link invariants 
are constructed after introducing a \cls\ operation. As in the case of  knots,
these invariants are expressed in term of multidimensional path  integrals along
the loops corresponding to the different components of the  link.

It is by now well known that Vassiliev invariants, which were originally
defined for knots, can be also defined for other objects as links, string
links, braids, tangles, etc.  These objects can be regarded as classes of
embeddings of one-dimensional objects in a three-dimensional space modulo some
kind of isotopy. Either from  quantum groups or from Chern-Simons gauge theory,
isotopy invariants can be constructed. These are formal Laurent polynomials in a
paramter $q$. The coefficients of the expansions of these invariants in power
series of $x$, being $q=\ex^x$, are Vassiliev invariants \cite{lin}. Perturbative
Chern-Simons gauge theory provides a way to construct geometrical or path
integral expressions for the resulting Vassiliev invariants. This fact should
hold for any of the objects quoted above. We will concentrate in this work in
the case of links but generalizations should be carried out for other cases.

The paper is organized as follows. Section 2 contains an elementary 
exposition of Chern-Simons quantum field theory, along with the definition of
Wilson-loop and Wilson-line operators and some nomenclature. In section 3 
we introduce the general structure of the perturbative expansion of 
a Wilson-line operator for two lines, 
and the definition of canonical bases. Section 4 contains a group-theoretical
result that, though simple, pervades the rest of this work. In section 5 we 
present the Master Equation, which is the key to the Theorem of 
Factorization; this theorem encodes the  
consequences of our having chosen a canonical basis to express the perturbative
expansion. In section 6 we define the opening and closing operation, and 
analyze the invariants so obtained. Explicit integral expressions for these
invariants are  presented in section 7 up to order four. These are computed for
all two-component links of no more than six crossings in section 8. Finally, in
section 9 we state our conclusions. Appendices A, B and
C,  contain details on our group-theoretical conventions and lists of the
polynomial invariants used in section 8.

\newpage
\setcounter{equation}{0}
\section{Chern-Simons theory}

In this section we will describe known results on Chern-Simons perturbation
theory. We do not attempt here to provide a derivation of these results. 
This can be studied in previous works \cite{numbers,gmm,alr}. What we will
do is to point out the salient features of the analysis of Chern-Simons gauge
theory in the framework of perturbation theory and to summarize the set of 
rules which comes out of that analysis. These rules are known as Feynman rules 
and they can be neatly described in terms of Feynman diagrams. The aim of this
section is therefore to provide the necessary framework so that the reader 
could write down the contribution to the vacuum expectation value of a product 
of Wilson-line operators at any order in perturbation theory. 

We will restrict ourselves to the case in which the three-dimensional manifold
is ${\RR^3}$ and the gauge group is a semi-simple compact Lie group $G$. Let 
$A$ be a $G$-connection. The action of the theory is the integral over
${\RR^3}$ of the Chern-Simons form:
\beq
S_k(A)={k\over 4\pi}\int_{\RRs^3} \tr (A\wedge dA + 
{2\over 3} A\wedge A\wedge A),
\label{action}
\eeq
where Tr denotes the trace in the fundamental representation of $G$ and $k$ is
a real parameter. As pointed out in \cite{gmm,alr,pert} there are three 
problems in the analysis of Chern-Simons gauge theory from the point of view of
perturbation theory. First, the theory based on the action (\ref{action})
has a gauge symmetry which has to be fixed. Invariance under gauge
transformations which are not connected to the identity implies certain
quantization conditions for the parameter $k$ \cite{esther}. From the point of 
view of perturbation theory this condition is not important and we will take
$g=\sqrt{4\pi/k}$ as the expansion parameter of the resulting perturbative 
power series. We will choose as gauge-fixing the Landau gauge considered in 
\cite{alr}. This gauge has the advantage of being covariant and free of 
infrared divergences.

The second problem that one has to face in the perturbative analysis of
Chern-Simons gauge theory is the presence of ultraviolet divergences. This
implies that the theory has to be regularized. As described in
\cite{gmm,alr} the theory does not have to be renormalized. Once the theory
is regularized and the regulator is removed each of the terms in the
perturbative expansion becomes finite. This means that no dimensionful 
parameter is needed to describe the theory at the quantum level. Different 
regularizations lead to different perturbative expansions which, however, are 
related by a redefinition of the parameter $k$. We will choose the 
regularization proposed in \cite{alr} and elaborated for higher loops in 
\cite{carmelo}. The salient feature of this regularization is that higher-loop 
contributions to the two and three-point functions account for a shift in 
$k$: $k\rightarrow k-C_A$, being $C_A$ the quadratic Casimir in the adjoint 
representation of $G$, so one can disregard  them from the perturbative 
expansion and take as expansion parameter $g=\sqrt{4\pi/(k-C_A)}$. 

Finally, one has to cure the intrinsic ambiguity appearing when products of
operators are evaluated at the same point. As shown in \cite{witten,gmm}
this problem can be solved without spoiling the topological nature of the
theory. However, in the process of fixing the ambiguity in this way 
one is forced to introduce an integer which will be identified with the framing
associated to some of the observables. To be more precise in the description 
of this effect we need first to introduce the types of operators which are 
present in Chern-Simons gauge theory. We will do this next and we postpone the
discussion on the role played by this third problem to the end of the next 
subsection.

The basic gauge invariant operators of Chern-Simons gauge theory which lead to
topological invariants are Wilson-loop operators. There are also graph 
operators \cite{wittengraphas} but these will not be considered in this 
paper. Wilson loops are labeled by a loop $C$ embedded in ${\RR^3}$ and a 
representation $R$ of the gauge group $G$, and correspond to the holonomy 
around the loop $C$ of the gauge connection $A$:
\beq
\W_R(C,G)=\tr\left[{\hbox{\rm P}}_R \exp \oint A \right].
\label{wilsonloop}
\eeq
In this equation ${\hbox{\rm P}}_R$ denotes path-ordered and the fact that
$A$ must be considered in the representation $R$: $A=A^a T_a^{(R)}$ being
$T_a^{(R)}$, $a=1,\dots,{\hbox{\rm dim}}(G)$, the generators of $G$ in the
representation $R$. In Chern-Simons theory one considers vacuum expectation
values of products of Wilson-line operators:
\bear
\langle \W_{R_1}(C_1,G)\W_{R_2}(C_2,G)\dots\W_{R_n}(C_n,G)\rangle =
{1\over Z_k}
\int [DA]& & \W_{R_1}(C_1,G)\W_{R_2}(C_2,G)\dots\ \nonumber\\ 
& &\W_{R_n}(C_n,G){\rm e}^{iS_k(A)}, 
\label{vev}
\eear
where $Z_k$ is the partition function:
\begin{equation}
Z_k=\int [DA]\, {\rm e}^{iS_k(A)}.
\label{parfun}
\end{equation}
As shown in \cite{witten} the quantity (\ref{vev}) is a link invariant 
associated to a colored $n$-component link whose $j$-component, $C_j$, carries 
the representation $R_j$, $j=1,\dots,n$.

In this paper we will be considering also operators which are not gauge
invariant but gauge covariant. If instead of a loop in (\ref{wilsonloop}) one
considers a line with fixed end points, the resulting operator, which we will
call Wilson-line operator, is gauge covariant. As shown in \cite{witten} these
operators also lead to interesting quantities from a topological point of view.
For specific choices of the three-manifold they are related to conformal blocks
\cite{witten,blocks} and are invariant under certain deformations of the lines
involved. We will label Wilson-line operators in the following way: 
\beq
\F_R(P,Q,L,G)_i{}^j=\left[{\hbox{\rm P}}_R \exp \int_P^Q A \right]_{i}^{\,\,j},
\label{wilsonline}
\eeq
where $P$ and $Q$ denote the two fixed end points of the line $L$, and $i$ and
$j$ run respectively over the representation $R$ and its conjugate $\bar R$.
The object of interest in Chern-Simons gauge theory is the product of $n$
Wilson-line operators,
\begin{eqnarray}
& &\langle
\F_{R_1}(P_1,Q_1,L_1,G)\F_{R_2}(P_2,Q_2,L_2,G)\dots\F_{R_n}
(P_n,Q_n,L_n,G)\rangle\nonumber\\
& &= {1\over Z_k} \int [DA]\, \F_{R_1}(P_1,Q_1,L_1,G)\F_{R_2}
(P_2,Q_2,L_2,G)\dots\F_{R_n}(P_n,Q_n,L_n,G)\,
{\rm e}^{iS_k(A)}. 
\label{vevwl} 
\end{eqnarray}
As shown in \cite{witten,blocks} for some specific choice of manifold and gauge
fixing this quantity is related to a conformal block on $S^2$ with $2n$ marked
points.

\subsection{Perturbative analysis and Feynman Rules}

The vacuum expectation values (\ref{vev}) and (\ref{vevwl}) can be calculated 
non-perturbatively by splitting the three manifold ($\RR^3$ in our case) into
two three manifolds with boundary, in which a WZW theory with sources is 
induced \cite{witten,eli,blocks,kaul}. We shall not follow that 
approach, but rather analyze the same objects within the framework of
perturbation theory. The main arguments have been explained before in 
\cite{numbers,factor,torus}, where the reader is referred to for
details. 

To fix ideas we shall outline the perturbative analysis of the \vev\ of the 
Wilson-loop operator (\ref{wilsonloop}). Simply stated, this \vev\ is 
evaluated as a formal power series in the 
variable $x=ig^2/2$ after rescaling the gauge field $A\rightarrow gA$ both in
the action (\ref{action}) and in the Wilson-loop operator. A useful
parametrization of that power series is \cite{numbers}:
\beq
\langle \W_R(C,G)  \rangle=d(R)\sum_{i=0}^{\infty}\sum_{j=1}^{d_i}
\alpha_i^{\,\,j}(C)\,r_{ij}(G)\,x^i,
\label{general}
\eeq
where the symbols $\alpha_i^{\,\,j}(C)$ are combinations of path integrals
of some kernels along the loop $C$ and over $\RR^3$, and the $r_{ij}$ are 
traces of products of generators of the Lie algebra associated with the gauge 
group $G$. The  index $i$ is called the ``order'' in perturbation theory, and 
$j$ labels independent contributions to a given order, being $d_i$ the number 
of these at order $i$. In (\ref{general}) $d(R)$ denotes the dimension of the
representation $R$.

\begin{figure}
\centerline{\hskip.4in \epsffile{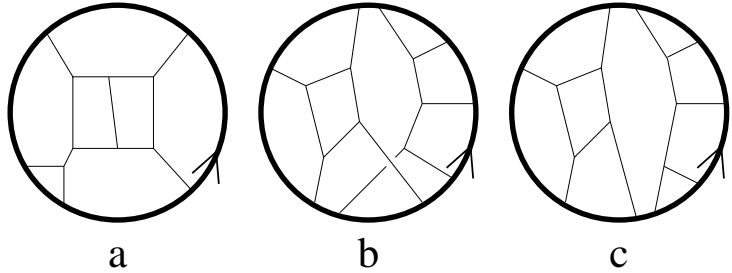}}
\caption{Some Feynman diagrams.}
\label{examples.eps}
\end{figure}

\begin{figure}
\centerline{\hskip.4in \epsffile{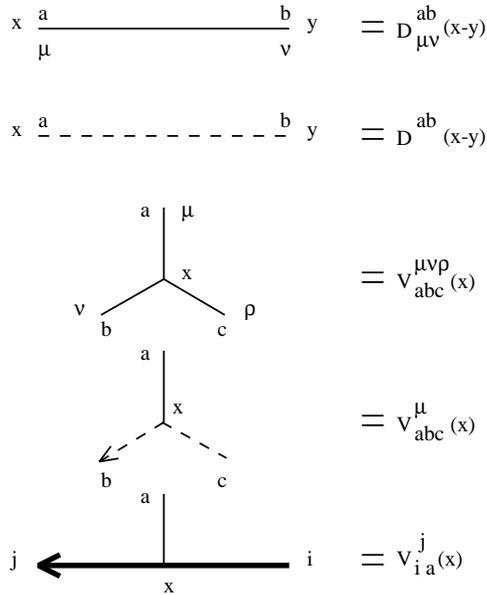}}
\caption{Feynman rules.}
\label{rules.eps}
\end{figure}

Each term in the expansion (\ref{general}) can be conveniently represented as 
a Feynman diagram like the ones depicted in Fig.~\ref{examples.eps}. These 
diagrams are constructed from the lines and vertices described in 
Fig.~\ref{rules.eps}. Each type of line corresponds to a kernel (or 
``propagator''), and each vertex to an integration; these correspondences are 
the Feynman rules. 

In Chern-Simons gauge theory the Feynman rules associated with the lines and
vertices of Fig.~\ref{rules.eps} are:
\bear
D_{\mu\nu}^{ab}(x-y)&=&{i\over 4\pi}
\epsilon_{\mu\rho\nu}{(x-y)^{\rho}\over |x-y|^3}\delta^{ab},\nonumber\\
D^{ab}(x-y)&=&{i\over 4\pi}{1\over |x-y|}\delta^{ab},\nonumber\\
V^{\mu\nu\rho}_{abc}(x)&=&-igf_{abc}\epsilon^{\mu\nu\rho}
\int_{\RRs^3}d^3x,
\nonumber \\
V_{abc}^{\mu}(x)&=& igf_{abc}\partial^{\mu}_x\int_{\RRs^3}d^3\omega, 
\nonumber\\
V_{i\,a}^{\,j}(x)&=& g\left(T_a^{(R)}\right)_i^{\,\,j}\int dx.
\label{kernels}
\eear

The argument $x$ in $V_{i\,a}^{\,j}(x)$ is a point on the Wilson line, and the
integration runs over a segment of the Wilson line limited by the two nearest 
insertions of the same vertex. The arrow on the last diagram of 
Fig.~\ref{rules.eps} indicates the orientation of the Wilson line. Since the 
structure constants of a semisimple Lie group can be chosen to be totally 
antisymmetric, there is no need to assign orientation to the internal 
three-vertices. The gauge lines need not be oriented either because the 
adjoint representation is real. With the help of the Feynman rules we can 
evaluate the \vev\ of any Wilson line operator to any order in perturbation 
theory once we have drawn all the diagrams that contribute at that order. 

The simplest non-trivial diagram would consist in a gauge propagator whith 
both endpoints attached to the Wilson loop; this corresponds to the term
$i=1$ in (\ref{general}). It is also the simplest way to introduce the
third difficulty mentioned above. Owing to the integration along the Wilson
loop, the two endpoints of the propagator will eventually get together (or
``collapse'' in the terminology of \cite{numbers}). Although the apparently
divergent integral is finite \cite{gmm}, it turns out that it is not 
invariant under small deformations of the Wilson loop, \ie it is not a 
topological invariant. The solution to this problem was proposed in 
\cite{witten} as attaching a ``framing'' to the Wilson loop; this framing is 
another loop defined by a small normal vector field along the Wilson 
loop. Attaching one of the endpoints of the propagator to the Wilson 
loop and the other to the framing the integration is well-defined and 
corresponds to computing the linking number of the framing around the Wilson 
loop. 

The only diagrams that perceive the existence of the framing are those with 
collapsible propagators; these are propagators whose endpoints may get together
without crossing over other propagators \cite{pert,numbers,factor}. It has 
been shown in those references that the contribution of the collapsible 
diagrams factorizes in an exponential, in total agreement with 
non-perturbative calculations \cite{witten}. The framing, however, is of no 
topological relevance and its contribution must therefore be discarded. We 
shall dispose of the framing by not including collapsible diagrams in the 
expansion (\ref{general}).

\subsection{Wilson Line Operator for Two Open Lines}

The aim of this work is to construct an approach to define numerical invariants
for links. We will describe in this paper the case of two-component links in
full detail. The generalization for links with an arbitrary number of 
components will be presented elsewhere. As described in the introduction  we 
shall analyze first the \vev\ of the product of two Wilson-line operators. 
Let us attach different irreducible representations $\lambda$ and $\mu$ of the 
gauge group $G$ to each line $L_1$ and $L_2$:
\beq
\F_{\lambda}(P_1, Q_1, L_1, G)_{i_1}^{\,\,j_1}
\F_{\mu}(P_2, Q_2, L_2, G)_{i_2}^{\,\,j_2}
={\rm{P}}_{\lambda}\exp\left\{ \int_{L_1}A\right\}_{i_1}^{\;\, j_1}
{\rm{P}}_{\mu}\exp\left\{ \int_{L_2}A
\right\}_{i_2}^{\;\, j_2}
\label{braid}
\eeq
where, say, $\rm{P}_\lambda$ has the usual meaning of path-ordering, and the
gauge field entering in the corresponding exponential is
$A(x)=A^a(x)(T_a^{(\lambda)})_i^{\,\,j}$. The Greek index in the
generators $\big(T_a^{(\lambda)}\big)_i^j$ designates the representation in
which they are defined; if it is $\lambda$, the indices $i,j$ run respectively
over the  representations $\lambda$ and its conjugate $\bar{\lambda}$. When not
strictly necessary these  indices, and other two coming from $\mu$, will not be
written explicitly in  the Wilson-loop operator in order not to clutter the
notation. For the same reason we shall omit all the arguments $P$, $Q$, and 
$G$, and let $\lambda$ or $\mu$ name both the representation of the  gauge group
defined in each line and, implicitly, the gauge group itself. 

Before we take a deeper look inside the general structure of
the perturbative expansion of the \vev\  of our operator,
\beq
\langle \F_{\lambda}(L_1)\F_{\mu}(L_2) \rangle,
\eeq 
let us introduce some vocabulary related to the full set of Feynman diagrams 
coming out of this expansion. These diagrams are built up with the 
propagators and three-vertices described in the previous subsection. For 
the case of two open lines these diagrams are trivalent graphs with two 
distinguished lines which will be called Wilson lines, carrying the  
representations $\lambda$ and $\mu$. The other lines correspond to
propagators  and are called internal lines. We shall refer to the set of
internal lines of  a given diagram as the Feynman graph. In order to classify
the different types  of diagrams, we will use the following definitions: 

\vskip .25cm

\noindent {\underbar {subdiagram}}: a specific subset of propagators in a
given diagram.

\vskip .25cm

\noindent {\underbar {connected (sub)diagram}}: we will say that a
(sub)diagram is a connected (sub)diagram if it is  possible to go
from one propagator  to another without ever having to go
through any of the two Wilson lines.

\vskip .25cm

\noindent {\underbar {disconnected diagram}}: the previous description is not
possible. The diagram will be made of some connected subdiagrams.

\vskip .25cm

\noindent {\underbar {non-overlapping subdiagrams}}: we say that two
subdiagrams are non-overlapping if starting, say by the upper points of
the two Wilson-lines, we can move along them meeting all the legs of one
subdiagram first, and all the legs of  the other in the second place. Here,
``legs'' means the propagators directly attached to the Wilson lines.

 \vskip .25cm

\noindent {\underbar {self-interaction subdiagram}}: a subdiagram living
only in either of the two Wilson lines.

 \vskip .25cm

\noindent {\underbar {interaction subdiagram}}: a subdiagram connecting the
two Wilson lines.

\vskip .25cm

\noindent {\underbar {standard diagram}}: a diagram either connected or
made of non-overlapping connected subdiagrams.

\vskip .25cm

\begin{figure}
\centerline{\hskip.4in \epsffile{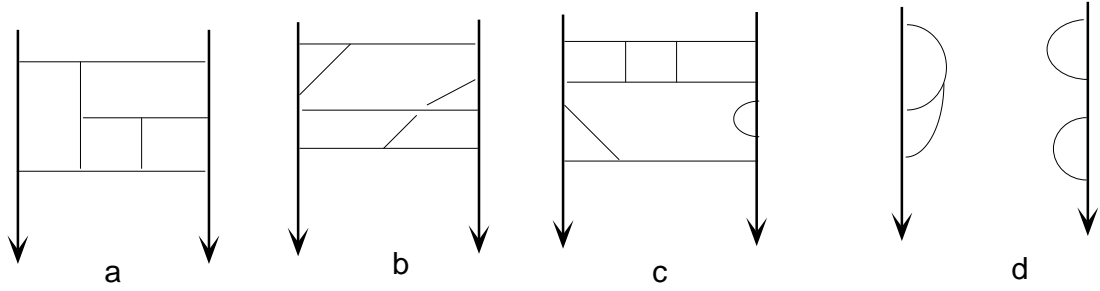}}
\caption{Examples of diagrams.}
\label{exdiag.eps}
\end{figure}

In Fig.~\ref{exdiag.eps} the diagram {\it a} is connected while the others 
are disconnected, containing subdiagrams which are connected. 
Diagram {\it a} contains only one connected subdiagram which
 coincides with itself.
In {\it c} and {\it d} the 
connected subdiagrams do not overlap, while in {\it b} they do. Diagrams {\it
a}, {\it b} and 
{\it c} contain  interaction and self-interaction subdiagrams while {\it d}
only self-interaction ones. All the subdiagrams in {\it a}, {\it c} and {\it d}
are non-overlapping.

\newpage
\setcounter{equation}{0}
\section{General structure of the perturbative expansion}

As pointed out in the introduction, our aim is to construct a framework to 
obtain finite type invariants for links. We will consider first the case of 
open two-component links and therefore we will analyze the perturbative series 
expansion of the operator (\ref{braid}). Recall that our main interest is to 
analyze two-component closed links, but their group factors do not satisfy a 
simple algebra and one must first analyze the case of open links. 

The strategy for analyzing open links is analogous to the approach described
in \cite{factor}. The perturbative series expansion we are studying now 
corresponds to the vacuum expectation value of the operator 
(\ref{braid}). All the possible Feynman diagrams which can be constructed 
from the Feynman rules enter in this expansion. Once we have excluded loop 
contributions from the two- and three-point functions and collapsible 
propagators, as was argued above, the perturbative expansion can be written as 
a generalization of (\ref{general}):
\begin{equation}
\langle \F_{\lambda}(L_1)\F_{\mu}(L_2) \rangle
=\sum_{i=0}^{\infty}\sum_{j=1}^{\dot D_i} 
A_i^{\,\,j}(L_1,L_2)R_{ij}(\lambda,\mu)x^i\,\, ,
\label{open}
\end{equation}
where $L_1$ and $L_2$ are the two open lines. The first line is coulored by 
the representation $\lambda$, and the second by the representation $\mu$. The 
factors $A_i^{\,\,j}$ and $R_{ij}$ in (\ref{open}) incorporate all the
dependence dictated from the Feynman  rules apart from the dependence on
$k$ which is contained in $x$. Of the two factors, $R_{ij}$ and
$A_i^{\,\,j}$, the first one contains all the group-theoretical dependence,
while the second all the geometrical dependence. The quantity $\dot D_i$
denotes   the number of independent group structures $R_{ij}$ which appear at
order $i$.

\begin{figure}
\centerline{\hskip.4in \epsffile{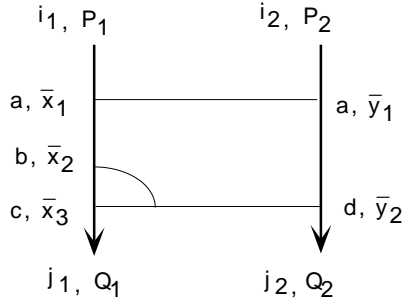}}
\caption{Example of group and geometrical factors.}
\label{exfactor.eps}
\end{figure}

Let us define properly the objects $R_{ij}$ and $A_i^{\,\,j}$. $R_{ij}$ is 
the product of two tensors. One comes from the product of generators 
in the first line, and the other from the product of generators in the second
line. Each  of these tensors has two indices corresponding to each endpoint of
the open line and a given number of indices in the adjoint representation of 
$G$; these are common to both tensors, and are contracted. Therefore, the 
tensor $R_{ij}$ has four indices: $i_1, j_1$  for $L_1$ and $i_2, j_2$ for 
$L_2$,
\beq
R_{ij}(\lambda,\mu) \rightarrow  \Big[  R_{ij} (\lambda,\mu)
\Big]_{i_1\,\,i_2}^{\,\,j_1\,\,j_2} \,\,.
\label{ogfactor}
\eeq
On the other hand, $A_i^{\,\,j}(L_1,L_2)$ is an integral over the two lines
which depends on four fixed points, the endpoints of the two Wilson lines, 
\beq
A_i^{\,\,j}  \rightarrow  A_i^{\,\,j}  (L_1,P_1,Q_1; L_2,P_2,Q_2),
\label{ogama}
\eeq
being $P_1$ and $Q_1$  respectively the endpoints of the first line, and  
$P_2$ and $Q_2$ those of the second line. As an example, the group 
and geometrical factors of the diagram in Fig.~\ref{exfactor.eps} are:
\beq
\big[ R_{ij} \big]_{i_1\,\,i_2}^{\,\,j_1\,\,j_2} = f^{bdc}
(T_a^{(\lambda)})_{i_1}^{\,\,m} (T_b^{(\lambda)})_{m}^{\,\,n} 
(T_c^{(\lambda)})_{n}^{\,\,j_1} 
(T_a^{(\mu)})_{i_2}^{\,\,p} (T_d^{(\mu)})_{p}^{\,\,j_2}
\label{sexemp}
\eeq

\begin{eqnarray}
 A_i^{\,\,j}(L_1,P_1,Q_1;L_2,P_2,Q_2) =& &{1\over 32}
\int^{Q_1} d\bar x_3^{\mu_3} 
      \int^{\bar x_3} d\bar x_{2}^{\mu_2} \int_{P_1}^{\bar x_2} d\bar
x_1^{\mu_1}
      \int^{Q_2} d\bar y_{2}^{\nu_2} \int_{P_2}^{\bar y_2} d\bar
y_1^{\nu_1}
\int_{\RRs^3} d^3\omega\, 
 \\ & & \epsilon^{\alpha \beta \gamma} \Delta_{\mu_1 \nu_1} (\bar x_1 - \bar y_1)
\nonumber \Delta_{\alpha\mu_2} (\bar x_2 -\omega) 
\Delta_{\mu_3 \beta} (\bar x_3 - \omega)\Delta_{ \gamma\nu_2} (\omega -
\bar y_2). 
\label{gamexemp}
\end{eqnarray}
where,
\begin{equation}
\Delta_{\mu \nu} (x - y) = {1\over \pi}\epsilon_{\mu\rho\nu}
{ (x-y)^\rho \over |x-y|^3}.
\label{elpropa}
\end{equation}

Notice that in defining group and geometrical factors the overal normalization
can be chosen arbitrarily. We will use a convention in which group factors are
taken to be the ones dictated by the Feynman rules without any additional
numerical factors. Once the group factor has been fixed and the expansion 
parameter $x$ extracted, the corresponding geometrical factor contains the 
rest of the ingredients dictated by the Feynman rules. This fixes completely 
the normalization ambiguity, but there is still some implicit dependence on 
the group-theoretical conventions used. The best convention to avoid 
ambiguities is to fix the values of the resulting primitive finite type 
invariant for a given link. For the case of knots it was noticed in 
\cite{numbers,torus} that there seems to exist a choice, at least up to order 
six, such that all the invariants are integer-valued. For the case of links, as
we will observe in sect. 7, it is not clear from the small amount of invariants
which we present if there exist a natural normalization such that all the 
primitive invariants are integer-valued.

\subsection{Canonical bases for two-component links}

In the general expansion (\ref{open}) there are many possible choices of
independent groups factors $R_{ij}$. Given all Feynman diagrams
contributing to a given order in perturbation theory, some of the resulting
group factors might be linear combinations of others due to the relations 
among the generators $T_a^{(R)}$ and the structure constants $f_{abc}$ of 
semi-simple groups. From a diagrammatic point of view these relations are the 
so-called STU and IHX relations \cite{drortopo}. A complete set of independent 
group factors at each order in perturbation theory will be called a ``basis'' 
of group factors \cite{numbers,factor}. 

The group factors entering (\ref{open}) are the elements of a given basis. Each
of these elements is represented by a Feynman diagram, of which we are only 
considering the group factor. This representation is of practical importance 
because it allows an index-free visual display of the elements of the basis. 
Besides, it simplifies considerably the tasks of calculating them for specific
gauge groups and of deciding when a given group factor is (in)dependent of
others. In this respect, we should have to indicate when, given a Feynman 
diagram, we are just considering its group factor and when we are only 
interested in its geometrical factor. In order to avoid a cumbersome notation 
we will not make this difference. It will be always clear from the context 
which case we are referring to.

Many choices of independent diagrams are possible. Each possible set of group
factors $R_{ij}$ represents a basis.  In order to study these bases we
establish first Proposition 1 which follows trivially from the 
group-theoretical properties of the group factors:
\vskip .5cm
{\bf Proposition 1:} {\it The group factor of a standard diagram $R_{ij}$ is 
the tensor product of the group factors of its subdiagrams.} 

For a choice of orientation for the Wilson lines as the one shown in 
Fig.~\ref{exfactor.eps}, the product is taken in the following way:
\beq
\bigg[R_{ij} \bigg]_{i_1\,\,i_2}^{\,\,j_1\,\,
j_2} = \bigg[ R_{ij}^{(1)} \bigg]_{i_1\,\,\,\,i_2}^{\,\,m_1\,\,m_2}
\bigg[
R_{ij}^{(2)} \bigg]_{m_1\,\,m_2}^{\,\,j_1\,\,\,\,j_2}
\label{product}
\eeq 
where $R_{ij}^{(1)}$ and $R_{ij}^{(2)}$ are the group factors of
two subdiagrams in $R_{ij}$. 

There are two simple  but far-reaching facts about the basis $R_{ij}$ which 
we summarize in  Propositions 2 and 3. 

\vskip0.2cm 
{\bf Proposition 2:} {\it It is always possible to choose a basis such that 
the $R_{ij}$ come from standard diagrams.} 

\vskip0.2cm {\bf Proposition 3:} {\it The $R_{ij}$ which are tensor products
can be chosen as tensor products of connected $R_{ij}$'s of lower orders}.

These propositions follow from a simple fact. Using STU
relations  it is always possible to trade in a disconnected diagram
overlapping  subdiagrams by connected diagrams and disconnected diagrams
containing  non-overlapping subdiagramas. A basis where these
propositions hold will be called canonical. One can easily see that a
canonical basis  shows the feature that a connected
$R_{ij}$ begets a whole family of group factors of higher orders, in
which it enters as a subdiagram.

The choice of a canonical basis allows us to classify the group factors into 
three different types:
\begin{equation}
R_{ij}(\lambda,\mu)=\left\{ r_{ij}(\lambda), r_{ij}(\mu), s_{ij}(\lambda,\mu)
, {\rm mixed}\right\} \,\, .
\label{clases}
\end{equation}

\begin{figure}
\centerline{\hskip.4in \epsffile{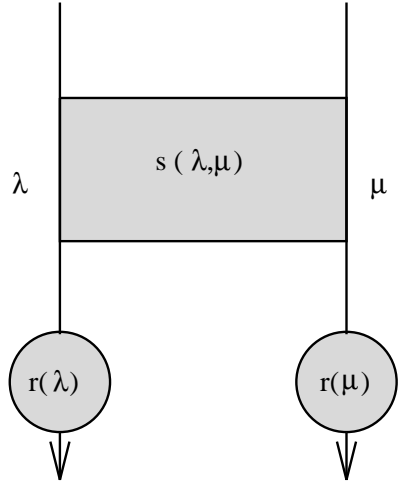}}
\caption{A general mixed element of a canonical basis.}
\label{general.eps}
\end{figure}

The first two sets $r_{ij}(\lambda)$ and $r_{ij}(\mu)$ are group factors
corresponding to diagrams made out of non-overlapping connected
self-interaction subdiagrams. Depending on their arguments ($\lambda$ or $\mu$)
they are attached to either one of the two Wilson lines. The third set 
($s_{ij}(\lambda,\mu)$) contains group factors which correspond to diagrams
made out of non-overlapping connected interaction subdiagrams. Finally, the
fourth set contains diagrams with both non-overlapping connected interaction 
and self-interaction subdiagrams. A general mixed diagram is shown in 
Fig.~\ref{general.eps}. 

\newpage
\setcounter{equation}{0}
\section{Group-theoretical considerations}

\begin{figure}
\centerline{\hskip.4in \epsffile{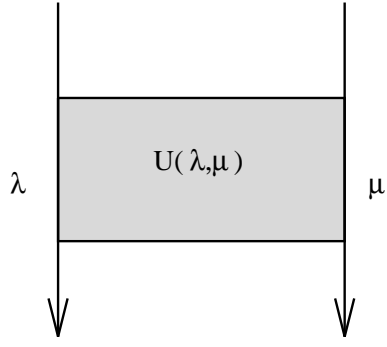}}
\caption{An invariant tensor $U(\lambda,\mu)$.}
\label{tensor.eps}
\end{figure}

Having adopted a canonical basis for our expansion (\ref{open}), we now show 
that the product (\ref{product}) of group factors corresponding to 
non-overlapping subdiagrams is commutative. This property is essential in order
to prove the Factorization Theorem below. Let us consider an arbitrary group 
factor as the one shown in Fig.~\ref{tensor.eps}, which represents a general 
connected interaction diagram; the dashed zone may include any Feynman 
graph. For this object we shall adopt a dual notation: it will be denoted by 
$U_{i_1\,\,i_2}^{\,\,j_1\,\,j_2}$ if we wish to indicate explicitly its 
representation indices; if not, it will be $U(\lambda,\mu)$. Before studying 
its properties let us recall some facts from group theory.

Let $i_1,j_1,k_1\ldots$ and $i_2, j_2, k_2 \ldots$ be indices for the unitary
irreducible representations (``irreps'') $\lambda$ and $\mu$ of the compact 
semisimple Lie group $G$. The product $\Psi_{\lambda}^{i_1}\Psi_{\mu}^{i_2}$ 
of two vectors corresponding to these two irreps decomposes in a Clebsch-Gordan
(CG) sum of vectors $\Psi_{\rho}^{i_3}$ where $\rho\subset\lambda\otimes\mu$:
\begin{equation}
\Psi_{\lambda}^{i_1}\Psi_{\mu}^{i_2}=\sum_{\rho\subset\lambda\otimes\mu}
\sum_{i_3=1}^{d(\rho)}
\left(\begin{array}{ccc} \lambda & \mu & \rho \\
i_1 & i_2 & i_3 \end{array}\right)\Psi_{\rho}^{i_3}\,\, .
\label{clebsch}
\end{equation}
This notation is abstract in the sense that the indices $i_1, i_2, $ etc., may
be, in a concrete case, composite indices. The quantity $d(\rho)$ represents 
the dimension of the irrep~$\rho$. 

The CG coefficients satisfy the completeness and orthogonality 
relations:
\begin{eqnarray}
& &\sum_{\rho\subset\lambda\otimes\mu}\sum_{i_3=1}^{d(\rho)}
\left(\begin{array}{ccc} \lambda & \mu & \rho \\
i_1 & i_2 & i_3 \end{array} \right)^{\ast} \left(\begin{array}{ccc} 
\lambda & \mu & \rho \\
j_1 & j_2 & i_3 \end{array}\right)=\delta_{i_1}^{\,\,j_1}\delta_{i_2}^{\,\,j_2}
\,\, ,\nonumber\\
& &\sum_{i_1=1}^{d(\lambda)}\sum_{i_2=1}^{d(\mu)}
\left( \begin{array}{ccc} \lambda & \mu & \rho \\
i_1 & i_2 & i_3 \end{array}  \right)^{\ast} 
\left( \begin{array}{ccc} \lambda & \mu & \rho' \\
i_1 & i_2 & i_3' \end{array}\right) =
\delta_{\rho'}^{\,\,\rho}\delta_{i_3'}^{i_3}     \,\,.
\label{complet}
\end{eqnarray}
From now on we shall assume that repeated representation indices 
(the Latin indices) are summed over. Now we insert the completeness relation 
in the tensor described in Fig.~\ref{tensor.eps}. One finds,
\begin{equation}
U_{i_1\,\,i_2}^{\,\,j_1\,\,j_2}=
\sum_{\rho,\rho'\subset\lambda\otimes\mu}U_{k_1\,\,k_2}^{\,\,l_1\,\,l_2}
\left(\begin{array}{ccc} \lambda & \mu & \rho \\
i_1 & i_2 & m \end{array} \right)^{\ast}\left( \begin{array}{ccc} 
\lambda & \mu & \rho \\
k_1 & k_2 & m \end{array}\right)
\left( \begin{array}{ccc} \lambda & \mu & \rho' \\
l_1 & l_2 & m' \end{array}\right)^{\ast}\left( \begin{array}{ccc} 
\lambda & \mu & \rho' \\
j_1 & j_2 & m' \end{array}\right)\,\,.
\label{decomp}
\end{equation}
In our investigation of the group factors associated to Feynman diagrams, the
tensors $U_{i_1\,\,j_1}^{\,\,i_2\,\,j_2}$ are constructed out of generators of
the Lie algebra $(T_a^{(R)})_i^{\,\,j}$ contracted with structure constants 
$f_{abc}$, Killing-Cartan metrics $\delta^{ab}$ or unit matrices 
$\delta_i^{\,\,j}$. A tensor such constructed is necessarily an invariant 
tensor \cite{cvit}. As the CG coefficients are also invariant tensors, we can
apply Schur's lemma to  
\begin{equation}
U_{k_1\,\,k_2}^{\,\,l_1\,\,l_2}\left(\begin{array}{ccc} 
\lambda & \mu & \rho \\
k_1 & k_2 & m \end{array} \right)
\left( \begin{array}{ccc} \lambda & \mu & \rho' \\
l_1 & l_2 & m' \end{array}\right)^{\ast}
\label{contract}
\end{equation}
since it is an invariant tensor with only two free indices corresponding to
the representations $\rho$ and $\rho'$. Given that both $\rho$ and 
$\rho'$ are irreps, Schur's lemma implies that 
\begin{equation}
U_{k_1\,\,k_2}^{\,\,l_1\,\,l_2}\left(\begin{array}{ccc} 
\lambda & \mu & \rho \\
k_1 & k_2 & m \end{array} \right)
\left( \begin{array}{ccc} \lambda & \mu & \rho' \\
l_1 & l_2 & m' \end{array}\right)^{\ast}
= U(\lambda\mu\rho)\delta_{\rho}^{\,\,\rho'}\delta_m^{\,\,m'},
\label{schur}
\end{equation}
where,
\begin{equation}
U(\lambda\mu\rho)
={1\over d(\rho)}
U_{k_1\,\,k_2}^{\,\,l_1\,\,l_2}\left(\begin{array}{ccc} 
\lambda & \mu & \rho \\
k_1 & k_2 & m \end{array} \right)
\left( \begin{array}{ccc} \lambda & \mu & \rho \\
l_1 & l_2 & m \end{array}\right)^{\ast}.
\label{schurdos}
\end{equation}
Inserting (\ref{schur}) in (\ref{decomp}) we 
arrive at a variant of the Wigner-Eckart theorem:
\begin{equation}
U_{i_1\,\,i_2}^{\,\,j_1\,\,j_2}=\sum_{\rho\subset\lambda\otimes\mu} U(\lambda
\mu\rho)
\left( \begin{array}{ccc} \lambda & \mu & \rho \\
i_1 & i_2 & m \end{array}\right)^{\ast} \left(\begin{array}{ccc} 
\lambda & \mu & \rho \\
j_1 & j_2 & m \end{array} \right) \,\,.
\label{wigner}
\end{equation}
The analogy with the Wigner-Eckart comes from the fact that the free indices 
factorize in each term of the CG sum in a structure independent of the 
tensor; all the information relative to the tensor is summarized in the 
scalars $U(\lambda\mu\rho)$.

\begin{figure}
\centerline{\hskip.4in \epsffile{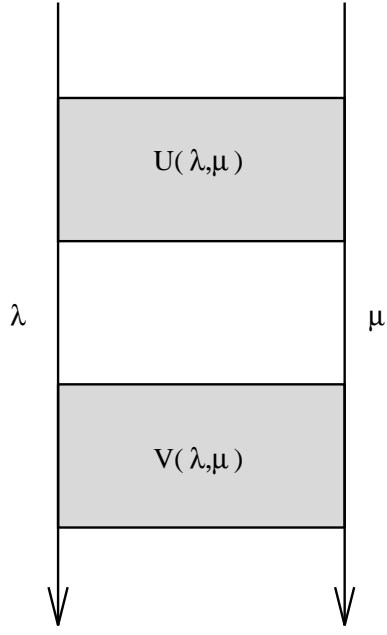}}
\caption{A more complex invariant tensor.}
\label{tensor2.eps}
\end{figure}

We turn now to a more complicated diagram, which consists of two connected
non-overlapping subdiagrams as depicted in Fig.~\ref{tensor2.eps}. It can 
be written as the product of two tensors of the type just considered:
\begin{equation}
U_{i_1\,\,i_2}^{\,\,j_1\,\,j_2}
V_{j_1\,\,j_2}^{\,\,p_1\,\,p_2}.
\label{prod}
\end{equation}
Applying our result (\ref{wigner}) and the relations (\ref{complet}), it 
follows that
\begin{equation}
U_{i_1\,\,i_2}^{\,\,j_1\,\,j_2}V_{j_1\,\,j_2}^{\,\,p_1\,\,p_2}=\!\!\!
\sum_{\rho\subset\lambda\otimes\mu}\!\! U(\lambda\mu\rho)V(\lambda\mu\rho)
\left( \begin{array}{ccc} \lambda & \mu & \rho \\
i_1 & i_2 & m \end{array}\right)^{\ast}  \left(\begin{array}{ccc} 
\lambda & \mu & \rho \\
p_1 & p_2 & m \end{array} \right).
\label{voila}
\end{equation}
We wish to emphasize that all free indices have been separated for each
irrep $\rho\subset\lambda\otimes\mu$ in a factor independent of the structure
of the tensors $U$ or $V$. The information relative to $U$ and $V$ is encoded 
in the scalars $U(\lambda\mu\rho)V(\lambda\mu\rho)$. From this formula it is 
clear that the product of connected non-overlapping subdiagrams is commutative 
in the sense that the order of the subdiagrams is irrelevant for the group 
factor of the diagram,
\begin{equation}
U_{i_1\,\,i_2}^{\,\,j_1\,\,j_2}V_{j_1\,\,j_2}^{\,\,p_1\,\,p_2}=
V_{i_1\,\,i_2}^{\,\,j_1\,\,j_2}U_{j_1\,\,j_2}^{\,\,p_1\,\,p_2}
\label{commut}
\end{equation}
This result can be generalized to diagrams with an arbitrary number 
of non-overlapping subdiagrams: the group factor of such a diagram does not
depend on the order of the subdiagrams. 

There is one further question regarding the tensors $U$ that will be relevant 
in what follows. Let us allow the gauge group to be the product of two compact
simple Lie groups: $G\times G'$. The corresponding irreps will be denoted by 
$\lambda\lambda'$. The generalization of the tensor $U(\lambda,\mu)$ is 
$U(\lambda\lambda',\mu\mu')$. If the diagram that represents
the tensor $U$ consists of several connected non-overlapping subdiagrams 
$U^{(p)}$ with $p=1,\ldots N$, it is easy to see that
\begin{equation}
U(\lambda\lambda',\mu\mu')=\prod_{p=1}^N\left(U^{(p)}(\lambda,\mu)I^{\lambda'}
I^{\mu'}+I^{\lambda}I^{\mu}U^{(p)}(\lambda',\mu')\right)\,\, ,
\label{ues}
\end{equation}
where $I^R$ denotes the $d(R)$-dimensional identity matrix.
Moreover, as a result of (\ref{commut}), whenever two of these $U^{(p)}$ 
corresponding to the same representations $\lambda$ and $\mu$ are 
multiplied, we need not care about the order in which they appear.

\newpage
\setcounter{equation}{0}
\section{The Master Equation}

In this section we shall demonstrate that the expansion (\ref{open}) is a 
product of three factors: two of them subsume all the information relative to 
each of the lines separately, while the third one encodes their 
``linkedness''. This will be made precise in the next subsection. 

A general element $R_{ij}$ of a canonical basis would look like the diagram in 
Fig.~\ref{general.eps}. The subdiagrams denoted by $r(\lambda)$, $r(\mu)$ 
and $s(\lambda,\mu)$ need not be connected; they may contain subdiagrams.  

Let a given $R_{ij}$ be composed of $p_{ij}$ subdiagrams of type $r$ on one 
Wilson line, $q_{ij}$ subdiagrams of type $r$ on the other Wilson line and 
$t_{ij}$ subdiagrams of type $s$; all these subdiagrams must be connected and
non-overlapping. We can write symbolically:
\begin{equation}
R_{ij}(\lambda,\mu)=\left\{ r_{ij}^{(p)}(\lambda), r_{ij}^{(q)}(\mu), 
s_{ij}^{(t)}(\lambda,\mu)\right\} \,\, ,
\label{composition}
\end{equation}
where $p=1,\ldots,p_{ij}$, $q=1,\ldots,q_{ij}$ and $t=1,\ldots,t_{ij}$. The 
indices $ij$ of the $r$'s and $s$ in (\ref{composition}) do not denote the 
order $i$ in perturbation theory or the element $j$ of the basis at each order 
as would be usual. Rather, they are a reminder that the subdiagrams are part of
the whole diagram $R_{ij}$. The order in perturbation theory of $r_{ij}^{(p)}$ 
or $s_{ij}^{(t)}$ will be denoted by $\x(ij,p)$ and $\x(ij,t)$ respectively. If
the gauge group is simple, it holds that
\begin{equation}
R_{ij}(\lambda,\mu)=\prod_{p=1}^{p_{ij}}r_{ij}^{(p)}(\lambda)
\prod_{q=1}^{q_{ij}}r_{ij}^{(q)}(\mu)
\prod_{t=1}^{t_{ij}}s_{ij}^{(t)}(\lambda,\mu)\,\, .
\label{simple}
\end{equation}
Let the gauge group be the product $G\times G'$ as by the end of the preceding 
section. The generalization of (\ref{simple}) is
\begin{eqnarray}
R_{ij}(\lambda\lambda',\mu\mu')&=&\prod_{p=1}^{p_{ij}}
\left(r_{ij}^{(p)}(\lambda)I^{\lambda'}+
I^{\lambda}r_{ij}^{(p)}(\lambda')\right)
\prod_{q=1}^{q_{ij}} \left(r_{ij}^{(q)}(\mu)I^{\mu'}
+I^{\mu}r_{ij}^{(q)}(\mu')\right)\\ & &
\prod_{t=1}^{t_{ij}}\left(s_{ij}^{(t)}(\lambda,\mu)I^{\lambda'}I^{\mu'}+
I^{\lambda}I^{\mu}s_{ij}^{(t)}(\lambda',\mu')\right)\nonumber
\,\, .
\label{nosimple}
\end{eqnarray}
The last ingredient we need is the following identity, which follows from the 
definition of the Wilson line operator:
\begin{equation}
\langle \F_{\lambda\lambda'}(L_1)\F_{\mu\mu'}(L_2)\rangle =
\langle \F_{\lambda}(L_1)\F_{\mu}(L_2)\rangle \langle \F_{\lambda'}(L_1)
\F_{\mu'}(L_2)\rangle \,\, ,
\label{last}
\end{equation}
Inserting the expansion (\ref{open}) in each of the factors
in (\ref{last}) we arrive at the Master Equation: 
\begin{eqnarray}
&&\sum_{i=0}^{\infty}\sum_{j=1}^{D_i}A_i^{\,\,j}(L_1,L_2)\prod_{p=1}^{p_{ij}}
\left(r_{ij}^{(p)}(\lambda)I^{\lambda'}x^{\x(ij,p)} +
I^{\lambda}r_{ij}^{(p)}(\lambda'){x'}^{\x(ij,p)}\right)\\ &&\times
\prod_{q=1}^{q_{ij}} \left(r_{ij}^{(q)}(\mu)I^{\mu'}x^{\x(ij,q)}
+I^{\mu}r_{ij}^{(q)}(\mu'){x'}^{\x(ij,q)}\right) \nonumber\\ &&\times 
\prod_{t=1}^{t_{ij}} \left(s_{ij}^{(t)}(\lambda,\mu)I^{\lambda'}I^{\mu'}
x^{\x(ij,t)}
+I^{\lambda}I^{\mu}s_{ij}^{(t)}(\lambda',\mu'){x'}^{\x(ij,t)}\right)\nonumber\\
&&=\left(\sum_{k=0}^{\infty}\sum_{l=1}^{D_k}A_k^{\,\,l}(L_1,L_2)
\prod_{p=1}^{p_{kl}}r_{kl}^{(p)}(\lambda)x^{\x(kl,p)}
\prod_{q=1}^{q_{kl}}r_{kl}^{(q)}(\mu)x^{\x(kl,q)}
\prod_{t=1}^{t_{kl}}s_{kl}^{(t)}(\lambda,\mu)x^{\x(kl,t)}\right) \nonumber\\
&&\times
\left(\sum_{m=0}^{\infty}\sum_{n=1}^{D_m}A_m^{\,\,n}(L_1,L_2)
\prod_{p=1}^{p_{mn}}r_{mn}^{(p)}(\lambda){x'}^{\x(mn,p)}
\prod_{q=1}^{q_{mn}}r_{mn}^{(q)}(\mu){x'}^{\x(mn,q)}
\prod_{t=1}^{t_{mn}}s_{mn}^{(t)}(\lambda,\mu){x'}^{\x(mn,t)}
\right) \nonumber \,\, .
\label{mastereq}\nonumber
\end{eqnarray}
When combined with the use of the same canonical basis in all the three 
expansions, this equation generates an infinite number of relations between 
geometric factors $A_i^{\,\,j}(L_1,L_2)$ at different orders.

\subsection{Factorization theorem}
In order to express the consequences of the master equation we need to 
introduce some notation. Let $A_i^{\,\,j}(L_1,L_2)$ be the geometric factor 
associated to a group factor $R_{ij}(\lambda,\mu)$ composed of connected 
non-overlapping subdiagrams of types $r$ and $s$ as in 
(\ref{composition}). Let 
\begin{eqnarray}
r_{ij}^{(1)}(\lambda)&=& r_{ij}^{(2)}(\lambda)=
\ldots r_{ij}^{(p_{ij}^{(1)})}(\lambda)\equiv r_{ij;1}(\lambda)\nonumber\\
r_{ij}^{(p_{ij}^{(1)}+1)}(\lambda)&=& r_{ij}^{(p_{ij}^{(2)}+2)}(\lambda)=
\ldots r_{ij}^{(p_{ij}^{(1)}+p_{ij}^{(2)})}(\lambda)\equiv r_{ij;2}(\lambda)
\label{classes}\\
&{\rm etc.}& \nonumber
\end{eqnarray}
and similar relations for the $r_{ij}^{(q)}(\mu)$ (changing also the $p$'s by
$q$'s) and for the $s_{ij}^{(t)}(\lambda,\mu)$ (changing the $p$'s by 
$t$'s). Equation~(\ref{classes}) is an enumeration of the possible identical 
subdiagrams ot type $r$ attached to the Wilson line that carries the
representation  $\lambda$, and similar enumerations hold for the subdiagrams
type $r$ in the  other Wilson lines and for the subdiagrams type $s$. The index
after the semicolon  labels the ``class'' of the subdiagram. For concreteness we
shall say that  there are P classes of subdiagrams $r(\lambda)$, Q classes of
subdiagrams  $r(\mu)$, and T classes of subdiagrams $s(\lambda,\mu)$. 

Let $\alpha_{i\,\,;u}^{\,\,j}(L_1)$ and 
$\alpha_{k\,\,;v}^{\,\,l}(L_2)$ be the geometric factor corresponding to 
$r_{ij;u}(\lambda)$ and $r_{kl;v}(\mu)$ respectively, and 
$\gamma_{m\,\,;w}^{\,\,n}(L_1,L_2)$ the geometric factor corresponding to 
$s_{mn;w}(\lambda,\mu)$. 

We can now express in a compact way the relations between different 
geometric factors stemming from the master equation:
\begin{equation}
A_i^{\,\,j}(L_1,L_2)=\prod_{u=1}^{\rm P}{1\over p_{ij}^{(u)}!}\left(
\alpha_{i\,\,;u}^{\,\,j}(L_1)\right)^{p_{ij}^{(u)}}
\prod_{v=1}^{\rm Q}{1\over q_{ij}^{(v)}!}\left(
\alpha_{i\,\,;v}^{\,\,j}(L_2)\right)^{q_{ij}^{(v)}}
\prod_{w=1}^{\rm T}{1\over t_{ij}^{(w)}!}\left(
\gamma_{i\,\,;w}^{\,\,j}(L_1,L_2)\right)^{t_{ij}^{(w)}}\,\, .
\label{factores}
\end{equation}
This equation is the generalization of the corresponding result for a single
closed loop \cite{factor}, and it holds only for canonical bases. Following the 
strategy described in \cite{factor} it is easy to conclude that 
Eq.~(\ref{factores}) implies that the vacuum expectation value of the
Wilson line operator (\ref{braid}) is the product of three exponentials:

\vspace{0.25truein}
{\bf Factorization Theorem}
\begin{equation}
\langle \F_{\lambda}(L_1)\F_{\mu}(L_2)\rangle =
\langle \F_{\lambda}(L_1)\rangle
\langle \F_{\mu}(L_2)\rangle \langle \LL_{\lambda,\mu}(L_1,L_2)\rangle,
\label{theofacto}
\end{equation}
where,
\begin{eqnarray}
\langle \F_{\lambda}(L_1)\rangle &=&
\exp\left\{ \sum_{i=0}^{\infty}\sum_{j=1}^{\hat{d}_i}
\alpha_i^{c\,\,j}(L_1)r_{ij}^c(\lambda)x^i \right\} \nonumber \\
\langle \F_{\mu}(L_2)\rangle &=&\exp\left\{ \sum_{i=0}^{\infty}
\sum_{j=1}^{\hat{d}_i}
\alpha_i^{c\,\,j}(L_2)r_{ij}^c(\mu)x^i \right\} \\
\langle \LL_{\lambda,\mu}(L_1,L_2)\rangle &=&\exp\left\{ 
\sum_{i=0}^{\infty}\sum_{j=1}^{\hat{\dot\delta}_i}
\gamma_i^{c\,\,j}(L_1,L_2)s_{ij}^c(\lambda,\mu)x^i \right\}\, .\nonumber 
\label{lastterm}
\end{eqnarray}
In this last equation we are restoring the original meaning of the 
symbols $r_{ij}$ and $\alpha_i^{\,\,j}$: the index $i$ denotes the order in
perturbation theory, and $j$ labels different independent contributions at a 
given order $i$. Similarly for the indices in $s_{ij}$ and $\gamma_i^{\,\,j}$.
We have added the superindex $c$ to denote that only the connected elements of
the canonical basis and their corresponding geometric factors must appear in
(\ref{theofacto}); the expansion of the exponentials generate  all the rest of
the diagrams. The numbers $\hat{d}_i$ and $\hat{\dot\delta}_i$  stand for the
number of independent group factors $r$ or $s$ at order $i$ corresponding to
connected diagrams. These group factors will be called primitive group
factors.  

\newpage
\setcounter{equation}{0}
\section{Opening and closing Wilson loops}

\hskip .25cm Our primary aim was to obtain the Vassiliev invariants for two
closed links. We begun studying the more general case of open links
because there we were able to find a complete set of relations defining the
algebra of the open geometric factors (eq. \ref{factores}). The property
of the $R_{ij}$ stated in Proposition 1, that is, the fact that the $R_{ij}$ 
of a diagram made of non-overlapping connected subdiagrams is the conmutative 
tensor product of the group factors of its subdiagrams, was fundamental to
obtain  them. But this property is lacking for closed links. 

At this point we have reached the central idea of this paper. Given that we
cannot apply directly the formalism of \cite{factor} to closed links, we must
``open'' them, apply the results of the preceding sections, ``close'' the 
lines back to the original link and see how things change. We shall clarify
this idea in the rest of this section. 

There are many inequivalent 
ways of closing an open link. Given a two-component open link it is always
possible to close its lines to end up with any arbitrary two-component 
closed link. Therefore we must define the relation between open and closed 
links more precisely if it is to be of any use. Let us suppose that we have a
two-component closed link $\LL$; the most natural prescription is the 
following:
\begin{enumerate} 
\item Select a point on each loop; call these two points $P$ and $P'$. 
\item Eliminate a small segment of each loop, starting on the selected 
point. We now have a two-component open link which we shall call 
$\dot{\LL}$. Let the end points be $P$, $P+\vec{\epsilon}$ for one line and 
$P'$, $P'+\vec{\epsilon'}$ for the other, where $\vec{\epsilon}$ and 
$\vec{\epsilon'}$ are elements of $\RR^3$ with very small components. The
endpoints carry also representation indices. Let these be $i$, $j$ for one
line and $i'$, $j'$ for the other. 
\item Apply the formalism described in the previous sections to $\dot{\LL}$.
\item Let $\vec{\epsilon}\rightarrow 0$, $\vec{\epsilon'}\rightarrow 0$ and
contract $i$, $j$ with a $\delta_i^{\,\,j}$ and $i'$, $j'$ with a 
$\delta_{i'}^{\,\,j'}$. 
\end{enumerate}

The first step is justified by the fact that the object of interest is the 
\vev\ of the Wilson line of a closed two-component link in the framework of 
Chern-Simons gauge theory, which we know to be a link invariant: 
\beq
\langle\calw _\lambda(C_1,G)\calw _\mu(C_2,G) \rangle= \langle \tr_{\lambda}
\left[{\hbox{\rm P}} \exp \oint_{C_1} A
\right] \tr_{\mu} \left[{\hbox{\rm P}} \exp \oint_{C_2} A
\right] \rangle,
\label{wilson}
\eeq 
where $C_1$ and $C_2$ are the two linked loops. In order to define the line
integrals in (\ref{wilson}) we have to parametrize each loop, and this already 
introduces a selected point on each of them. These selected points can be 
chosen to coincide with $P$ and $P'$ in step 1.

\begin{figure}
\centerline{\hskip.4in \epsffile{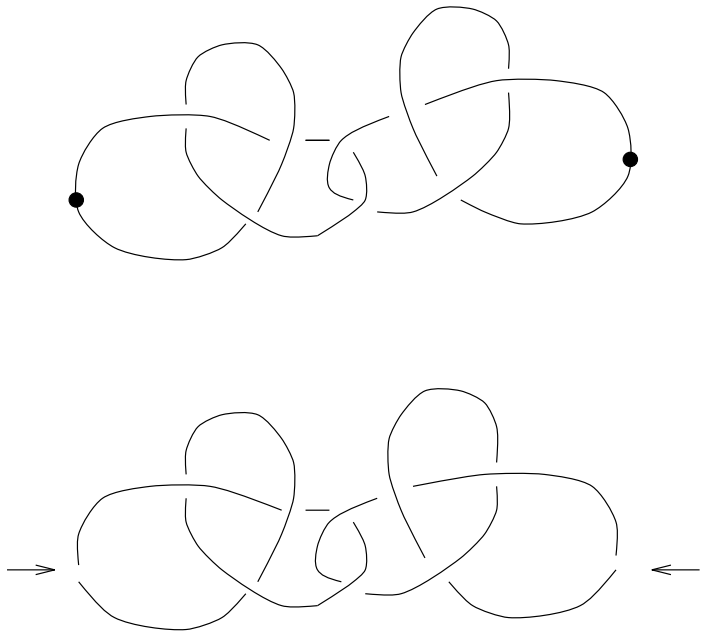}}
\caption{Opening a closed link. The arrows indicate the cuts.}
\label{link.eps}
\end{figure}

Steps 1 and 2 are roughly represented in Fig.~\ref{link.eps}. The open link 
$\dot{\LL}$ is an auxiliary entity which enables us to extract the geometric 
factors $\gamma_i^{\,\,j}$ described in the previous section. Step 2 creates a
$\dot{\LL}$ as similar to $\LL$ as possible. This step is similar to a
point-splitting  regularization in that the small vectors $\vec{\epsilon}$ and
$\vec{\epsilon'}$ must be small enough to avoid ``forgetting'' the original
shape of the closed  link. 

Step 4 defines the \cls\ operation. This operation should be understood as a 
generalized trace which operates both on the representation indices $i$, $j$, 
$i'$, $j'$ and on the coordinates of the endpoints of the open lines. 

The $\gamma_i^{\,\,j}$ of the open link are not topological invariants, but 
rather depend on the shape of the open link even if we let the four endpoints 
fixed. They also depend on $\vec{\epsilon}$ and $\vec{\epsilon'}$ through their
dependence on the endpoints of the open lines. The relevance of \cls\ is that
the $\gamma_i^{\,\,j}$ become topological invariants of the link in the limit 
$\vec{\epsilon}\rightarrow 0$, $\vec{\epsilon'}\rightarrow 0$.

These four steps describe our approach to defining numerical invariants of 
closed links. In what follows we shall analyze the effect of the \cls\ 
operation on the geometric factors $\gamma_i^{\,\,j}$. It will always be 
understood that the open links have been generated from closed links as 
described above. 

The perturbative expansion of (\ref{wilson}) can be written as:
\beq
\langle \calw _\lambda(C_1,G)\calw _\mu(C_2,G)  \rangle =
d(\lambda)d(\mu) \sum\limits_{i=0}^{\infty}
\sum\limits_{j=1}^{ D_i}  A_i^{\,\,j}(C_1,C_2)  R_{ij}(\lambda,\mu) x^i
\label{closed}
\eeq
The notation is the same as in (\ref{open}), except that now we are dealing 
with closed links. From now on, to distinguish between objects pertaining to 
open lines from objects pertaining to loops, we shall denote the former with a 
{\it dot} above them. The $R_{ij}$ are no longer tensors but numbers, because 
now we have to take the trace over the Wilson lines, and the $A_i^{\,\,j}$ 
represent integrals over two closed loops which can be shown not to depend on
the parametrization (so they do not depend on the choice of $P$ and $P'$ 
described in Step 1 above). 

\begin{figure}
\centerline{\hskip.4in \epsffile{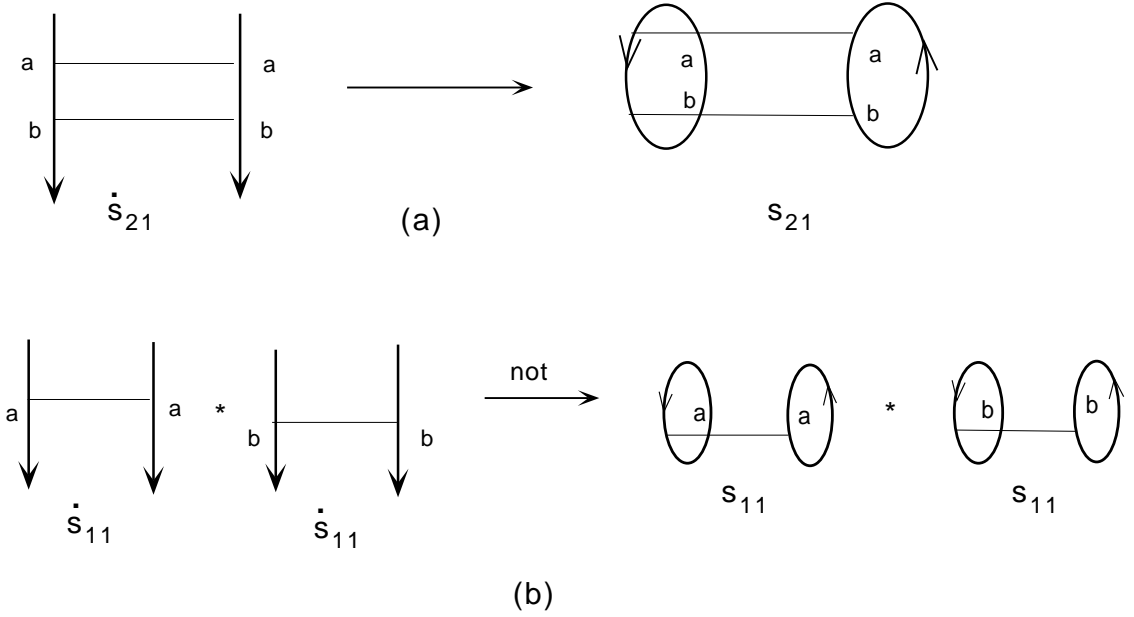}}
\caption{Closed group factors do not factorize.}
\label{closing.eps}
\end{figure}

Of course the dimension $D_i$ will be in general different from 
$\dot D_i$ in (\ref{open}). More precisely, it will be lower because taking
the traces nullifies some group factors or makes some of them identical. Also, 
if some group factors were dependent when the Wilson line was open the trace 
will never render them independent. The \cls\ operation is such that it 
transforms the $\dr {ij}$ and $\dot A_i^{\,\,j}$ of open lines in the 
corresponding factors of two oriented loops. Formally:
\begin{eqnarray} & & \dr {ij} \;  \longrightarrow R_{ij}, \nonumber \\ & &
\da i^{\,\,j} \;  \longrightarrow A_i^{\,\,j}. 
\label{closop}
\end{eqnarray} 
We will denote this operation by $\calc$; it is
pictorically represented for an example in 
Fig.~\ref{closing.eps}~(a). Although we are applying
$\calc$ only to the independent group and geometrical factors, it
obviously affects the whole set of Feynman diagrams. With respect to the
group factors, the \cls\ operation consists of taking the traces over each and 
all tensors corresponding to both Wilson lines.
\beq 
R_{ij}(G) = \calc \big[  \big( \dr {ij} \big)_{i_1\,\,i_2}^{\,\,j_1\,\,j_2}
\big] := \big( \dr {ij} \big)_{i_1\,\,i_2}^{\,\,i_1\,\,i_2}
\label{closrs}
\eeq
For the diagram in  Fig.~\ref{closing.eps}~(a) we have:
\beq
\big(T_a^{(\lambda)}T_b^{(\lambda)}\big)_{i_1}^{\,\,j_1}
 \big(T_b^{(\mu)} T_a^{(\mu)}\big)_{i_2}^{\,\,j_2}
\rightarrow  \tr (T_a^{(\lambda)}T_b^{(\lambda)}) \tr
(T_b^{(\lambda)} T_a^{(\lambda)})
\label{exclgroup}
\eeq As for the geometrical factors, \cls\ means identifying the endpoints of 
each line, so that we finish with integrals over closed loops:
\beq
A_i^{\,\,j}(C_1,C_2)=
\calc\big[\dot A_i^{\,\,j}\big(L_1,P_1,Q_1; L_2,P_2,Q_2\big)
\big]
\label{closas}
\eeq 
For example in Fig.~\ref{closing.eps}~(a) one has:
\beq
\int^{Q_1}\! d\bar x_{2}\! \int_{P_1}^{\bar x_2}\! d\bar x_1\!
\int^{Q_2}\! d\bar y_{2\!} \int_{P_2}^{\bar y_2}\! d\bar y_1 f(\bar x_1,\bar
x_2,\bar y_1,\bar y_2)
\longrightarrow \oint\! d\bar x_{2}\! \int^{\bar x_2}\! d\bar x_1\! \oint\!
d\bar y_{2}\! \int^{\bar y_2}\! d\bar y_1 f(\bar x_1,\bar x_2,\bar y_1,
\bar y_2)
\label{exclgeom}
\eeq 
where $f(\bar x_1,\bar x_2,\bar y_1,\bar y_2)$ is the corresponding integrand.

Let us analyze  the example shown in Fig.~\ref{closing.eps} to observe that 
Proposition 1, which was introduced for open Wilson lines, does not hold for 
Wilson loops. For open Wilson lines we have:
\begin{eqnarray}
\big[ \dot s_{21} \big]_{i_1 \,\,i_2}^{\,\,j_1\,\,j_2} &=& 
(T_a^{(\lambda)})_{i_1}^{\,\,m}(T_b^{(\lambda)})_{m}^{\,\,j_1} 
(T_a^{(\mu)})_{i_2}^{\,\,n}
(T_b^{(\mu)})_{n}^{\,\,j_2} =                    
(T_a^{(\lambda)})_{i_1}^{\,\,m}(T_a^{(\mu)})_{i_2}^{\,\,n}
(T_b^{(\lambda)})_{m}^{\,\,j_1} (T_b^{(\mu)})_{n}^{\,\,j_2}\nonumber \\ 
&=& \big[ \dot s_{11} \big]_{i_1\,\,i_2}^{\,\,m\,\,n} 
\big[ \dot s_{11} \big]_{m\,\,n}^{\,\,j_1\,\,j_2} 
\label{notpro}
\end{eqnarray}
but for loops:
\begin{eqnarray} 
\big[ s_{21} \big]_{i_1\,\,i_2}^{\,\,i_1\,\,i_2} &=&
\tr (T_b^{(\lambda)} T_a^{(\lambda)}) \tr (T_a^{(\mu)} T_b^{(\mu)})  \ne  
\tr(T_a^{(\lambda)}) \tr(T_a^{(\mu)})  \tr (T_b^{(\lambda)})
\tr(T_b^{(\lambda)})\nonumber\\ 
&=& \big[ s_{11} \big]_{i_1\,\,i_2}^{\,\,i_1\,\,i_2}
\big[ s_{11} \big]_{j_1\,\,j_2}^{\,\,j_1\,\,j_2}  
\label{noclose}
\end{eqnarray}
Our next task is to apply this \cls\ operation to the results of the previous 
chapter and find out the relations holding in the case of loops.

\subsection{Factorizing out the  part corresponding to knots}

The aim of this subsection is to show that the factorization of the 
disconnected contributions that was obtained for Wilson lines in 
(\ref{theofacto}) can be partially  extended to the case of loops. Owing to the
different algebraic structure of the group factors, a total factorization 
cannot be achieved. We shall show, however, that the individual contributions 
from each of the Wilson loops do factorize. Our starting point is the 
factorization theorem contained in eq.~(\ref{theofacto}). In that equation the 
terms $\langle \F_{\lambda}(L_1) \rangle$ and $\langle \F_{\mu}(L_2) \rangle$ 
are diagonal in the group indices and therefore proportional to the identity
matrix in their respective representations $\lambda$ and $\mu$. This means that
after closing, (\ref{theofacto}) becomes:
\beq
\langle \calw_\lambda (C_1,G) \calw_\mu (C_2,G)\rangle = 
\langle \calw_\lambda (C_1,G)
\rangle\langle \calw_\mu (C_2,G)
\rangle \langle \calz_{\lambda\mu} (C_1,C_2,G) \rangle
\label{knotfact}
\eeq
where $\calw_\lambda (C_1,G)$ is the Wilson-loop operator in
(\ref{wilsonloop}) and $\langle \calz_{\lambda\mu} (C_1,C_2,G) \rangle$
is the quantity which results of applying the \cls\ operation to 
$\langle \LL_{\lambda\mu} (L_1,L_2,G) \rangle$ in (\ref{theofacto}).
The loops resulting from the closing of the  lines $L_1$ and $L_2$ have been
denoted by $C_1$ and $C_2$ respectively. This last part, which we have been
calling interaction part, is the pure link contribution of the perturbative
series. The meaning of equation (\ref{knotfact}) is, on the one hand, that the
contribution from the knot invariant associated to each loop factorizes
from the full vacuum expectation value (\ref{vev}), on the other hand, that 
one has well defined intrinsic numerical link invariants. These link invariants
are indeed the geometrical factors originated from $\gamma_i^{c\,j}(L_1,L_2)$ 
in (\ref{theofacto})
after performing the \cls\ operation. Our next task is to describe the
properties of these link invariants.

\subsection{Numerical link invariants}

Let us begin rewriting the last term in (\ref{lastterm}) with our new notation
in which quantities for open Wilson lines are denoted with dots on top:
\beq
\langle \LL_{\lambda,\mu}(L_1,L_2)\rangle =\exp\left\{ 
\sum_{i=0}^{\infty}\sum_{j=1}^{\hat{\dot\delta}_i}
\dot\gamma_i^{c\,\,j}(L_1,L_2) \,\dot s_{ij}^c(\lambda,\mu)x^i \right\}.
\label{expint}
\eeq
Recall that the sum is over the primitive elements of the algebra. 
The \cls\ operation does not preserve the product of the group factors present
in (\ref{expint}):
\beq
\calc \big[\ds {ij} \ds {kl} \big] \ne \calc \big[\ds {ij} \big] \calc
\big[ \ds {kl} \big],
\label{nopres}
\eeq
where $\ds {ij}$ and $ \ds {kl}$ might or might not be primitive.
This implies that the result of applying the \cls\ operation to
(\ref{expint}) has to be analyzed order by order. Notice also that in this
operation we are losing information and therefore one expects a lower number
of independent group factors. As we will observe below, it is still possible
to have a notion of primitiveness for the resulting link invariants.

We introduce for the quantity
$\langle \calz_{\lambda\mu} (C_1,C_2,G) \rangle$ in (\ref{knotfact})  the
following perturbative series expansion: 
\beq
\langle \calz_{\lambda,\mu}(C_1,C_2,G)\rangle =\left\{ 
\sum_{i=0}^{\infty}\sum_{j=1}^{{\delta}_i}
\gamma_i^{\,\,j}(C_1,C_2) \, s_{ij}(\lambda,\mu,G) x^i \right\}.
\label{expintdos}
\eeq
where $\delta_i$ is the number of independent closed group factors.
Notice that we have restored the dependence of $s_{ij}$ on the group $G$.

Closed geometrical factors will not simply be the closure of open ones. If 
two or more group factors become proportional after the closing operation, we 
will choose one of them as the independent one; the closed geometric factor will 
be the sum of these geometric factors with the appropriate factors. Once the 
independent group factors $s_{ij}(\lambda,\mu,G)$ have been chosen, their 
corresponding geometrical factors are linear combinations of the open ones.

Using the relations
(\ref{factores}) of the algebra of the open geometrical factors, we can find
out the algebra of the new invariants for  two-component links.
Notice that in general we may not have pure diagonal relations like in
(\ref{factores}), due to the presence of the linear relations of the type
relating link invariants to closed geometrical factors.
It might exists, however, a suitable choice of basis in which most of the
relations would be diagonal. This happens at least up to order four. 

The invariance of the geometrical factors $\gamma_i^{\,\, j}$ follows from
 the general properties of Chern-Simons gauge theory. Being the vacuum
expectation value invariant, each term in the perturbative power series
expansion is also invariant. If the contributions appearing in each term
are organized in terms of the independent group factors, the coefficient of
each independent group factor is invariant. These coefficients are precisely
the geometrical factors $\gamma_i^{\,\, j}$. A simple
application of the theorems in
\cite{birlin,birman} shows that these geometrical factors are indeed
Vassiliev invariants or numerical invariants of finite type.
The arguments are the same as the ones presented in \cite{numbers}.

In the next
chapter we wil construct the link invariants introduced in this section up to
order four, and we will compute them for all two-component links of no more than
six crossings.

\newpage
\setcounter{equation}{0}
\section{Explicit results up to order four}

The aim of this section is to apply the results of the previous sections to
obtain explicit expressions for the numerical link invariants
$\gamma_{i}^{\,j}(C_1,C_2)$ in (\ref{expintdos})
up to order four in perturbation theory. 
We will begin analyzing first  the situation corresponding to
open lines. Then we will carry out the closing operation to achieve our goal.

\begin{figure}
\centerline{\hskip.4in \epsffile{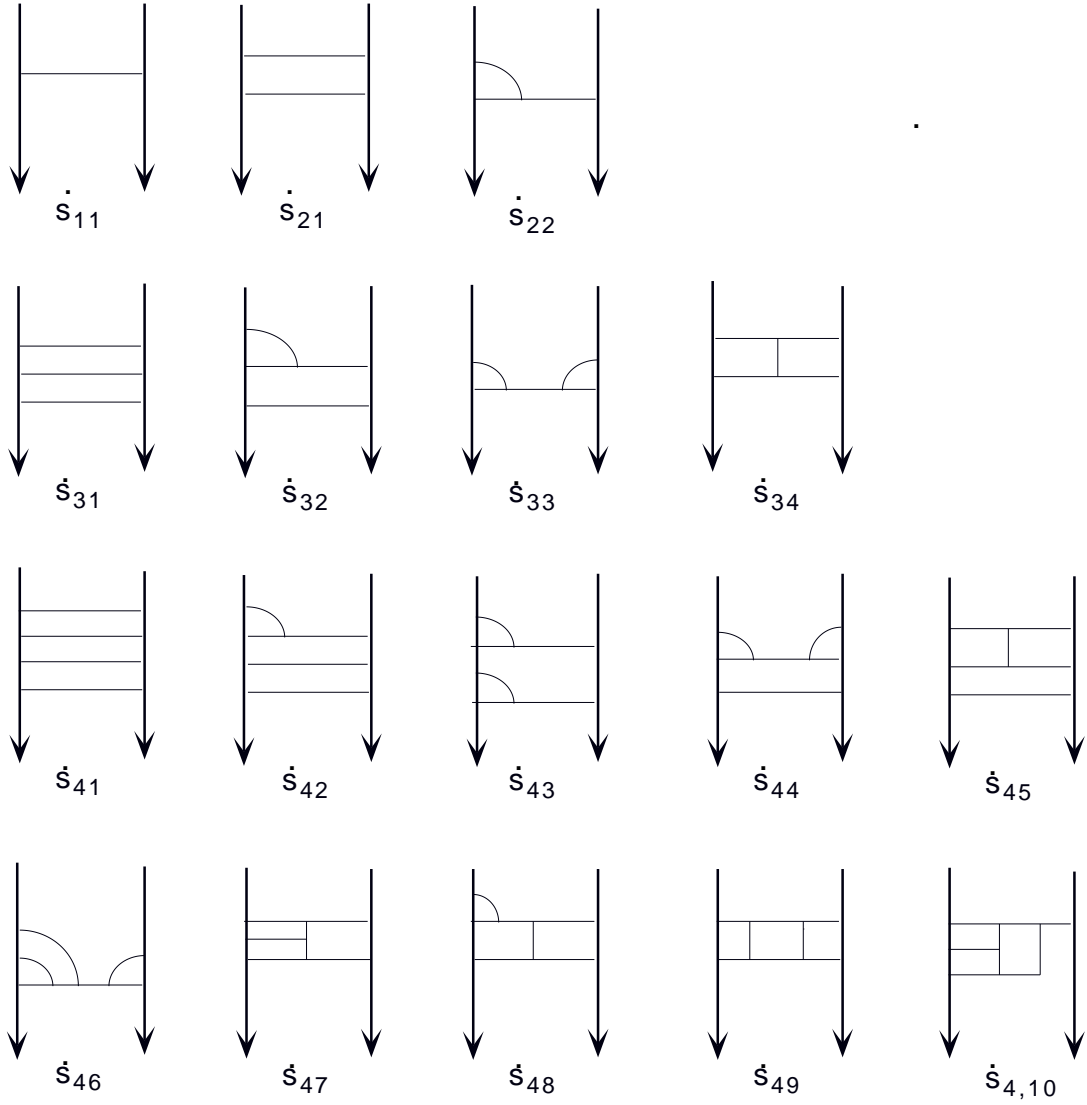}}
\caption{Basis up to order four.}
\label{basem.eps}
\end{figure}

Our first task is to choose a basis
for the $\dot s_{ij}(\lambda,\mu,\otimes_k G_k)$ in (\ref{open}), with
$G_k$ simple. Among them we will select the primitive group factors, which will
be the ones entering (\ref{expint}). The procedure is the following, write all
possible group factors at each order, use STU and IHX to find relations among
them, and extract a set of independent ones. Up to order four this computation
is rather simple but it gets complicated as the order is increased. The result
is depicted in Fig.~\ref{basem.eps}. Among all those group factors there are 
only one primitive group factor at first and second orders, $\dot s_{11}$ and 
$\dot s_{22}$, two at order three, $\dot s_{33}$ and $\dot s_{34}$, and five at
order four, $\dot s_{46}$, $\dot s_{47}$, $\dot s_{48}$, $\dot s_{49}$ and 
$\dot s_{4,10}$. Using this analysis we can then use (\ref{factores}) to write 
the $\dga {i}^{\,j}$ in terms of the  primitive elements of the  associated 
geometrical factors.  One finds

\begin{eqnarray}
& & \dga {n}^{\,1} = {(\dga {1}^{\,1})^n \over n!}, \;\; n = 1 \cdots
4, \nonumber \\ & & \dga {n}^{\,2} = {(\dga {1}^{\,1})^{n-2} \over (n-2)!} 
\dga {2}^{\,2}, \;\; n = 3
\, , 4, \nonumber \\ & & \dga {4}^{\,3} = {(\dga {2}^{\,2})^2 \over 2!} 
\; , \; \; \dga {4}^{\,4} = \dga {1}^{\,1} \dga {3}^{\,3} \; , \;\;
\dga {4}^{\,5} = \dga {1}^{\,1} \dga {3}^{\,4} \; , \;\;
\label{primi}
\end{eqnarray}

Notice that the primitive elements are the geometrical
factors associated to the connected and independent $\ds {ij}$. The fact
that $G=\otimes_k G_k$ is semi-simple affects directly the dimension $\dot
\delta_i$. If we were working with a simple group not all the diagrams pictured
in Fig.~\ref{basem.eps} would be independent. For example,  $\dot s_{43}$ and
$\dot s_{44}$ would be proportional. In general, the dimension $\dot \delta_i$ 
would be lower for the  case of simple groups. This has implications for the 
link invariants obtained after closing and therefore one can affirm that, as 
in the case of knots, simple Lie algebras are not enough to determine all 
invariants.  As expected from the factorization theorem, using relations
(\ref{primi}) we can write the perturbative series of $\langle
\LL_{\lambda,\mu}(L_1,L_2)\rangle$ up to order four in the  following form:
\begin{eqnarray}
& & 1 + x \dga {1}^{\,\,1} \ds {11} + x^2 \bigg[ {(\dga {1}^{\,\,1})^2 \over
2!} \ds {11}^2 + \dga {2}^{\,\,2} \ds {22} \bigg]
 + x^3 \bigg[ {(\dga {1}^{\,\,1})^3 \over 3!} \ds {11}^3 + \dga {1}^{\,\,1}
  \dga {2}^{\,\,2} \ds {11} \ds {22} +
 \dga {3}^{\,\,3} \ds {33} + \dga {3}^{\,\,4} \ds {34} \bigg]  \\
 & & + x^4 \bigg[ {(\dga {1}^{\,\,1})^4 \over 4!} \ds {11}^4 + 
{(\dga {1}^{\,\,1})^2 \over
2!}\dga {2}^{\,\,2} \ds {11}^2 \ds {22} + {(\dga {2}^{\,\,2})^2 \over 2!} \ds
{22}^2 + \dga {1}^{\,\,1} 
\dga {3}^{\,\,3} \ds {11} \ds {33} +
\dga {1}^{\,\,1}  \dga {3}^{\,\,4} \ds {11} \ds {34} + 
\sum\limits_{j=6}^{10} \dga {4}^{\,\,j}
\ds {4j} \bigg] + O(x^5). \nonumber
\label{serfour}
\end{eqnarray}
Our next task is to apply the closing operation to this expression.
Recall that the open group factors are depicted in Fig.~\ref{basem.eps}. We 
will carry out this analysis order by order, describing in detail the choices 
made.

\vskip .25cm
\noindent {\it i)} Order one. In this case one finds that the resulting closed
group factor vanishes unless we extend the scope of Lie groups under
consideration beyond the semi-simple case. Indeed, one has:
\beq
s_{11} = \calc [\ds {11}] \ne 0 \Leftrightarrow G = \otimes_k G_k
\otimes U(1)
\label{sone}
\eeq 
Nothing is lost restricting ourselves to the semi-simple case since the
geometrical factor $\gamma_1^{\,1}$, which is multiplying
$s_{11}$ in the perturbative series, appears at
higher orders. We will maintain the gauge group $G$ to be semi-simple and
therefore $\delta_1 = 0$. For these groups there is not linear term in
the perturbative series expansion (\ref{serfour}). Recall that for knots linear
terms appear only if a non-trivial framing is attached to the knot. Thus
according to the factorization formula (\ref{knotfact}), if the group is
semi-simple and the two components of the link are taken in the standar framing
($n=0$), no linear term in $x$ would appear in the expansion of any polynomial
link invariant.

\vskip .25cm
\noindent {\it ii)} Order two. In this case one finds,
\begin{eqnarray}
& & \calc [\ds {22}] = 0, \nonumber \\ & & s_{21} =
\calc [\ds {11}\ds {11}],
\label{stwo}
\end{eqnarray}
and therefore one has $\delta_2=1$. Notice that in this
case $\calc [\ds {22}] = 0$ even if the gauge group had a $U(1)$ factor.
In (\ref{stwo}) and in similar equations below the product of group factors
inside the squared brackets has to be understood as a tensor product.

\vskip .25cm
\noindent {\it iii)} Order three. In this case one finds the following set of
relations:
\begin{eqnarray}
&& s_{31} = \calc[\ds {11}\ds {11}\ds {11}], \nonumber \\ & & s_{32} =
\calc [\ds {11} \ds {22}], \nonumber \\ & & \calc [\ds {33}] = 0,
\;\;\;\;\;\calc [\ds {34}] = 2 s_{32}, 
\label{sthree}
\end{eqnarray}
so that $\delta_3=2$.

\vskip .25cm
\noindent {\it iv)} Order four. At this order the number of relations 
increases notably:
\begin{eqnarray}
& & s_{41} = \calc [\ds {11}\ds {11}\ds {11}\ds {11}], \nonumber \\ 
& & s_{42} = \calc [\ds {11}\ds {11} \ds {22}], \nonumber \\ 
& & s_{43} = \calc [\ds {22}\ds {22}], \nonumber \\ 
& & \calc [\ds {11}\ds {33}]
= -s_{43}, \;\;\;\;\; \calc [\ds {11}\ds {34}] =  s_{42},\nonumber 
\\ & & \calc [\ds {46}] = 0, \;\;\;\;\;\calc [\ds {47}] =  s_{43}, \nonumber 
\\ & & \calc [\ds {48}] = 2s_{43}, \;\;\;\;\; \calc[\ds {49}] = 4 s_{43},
\;\;\;\;\; \calc [\ds {4,10}] = 0.  \label{sfour}
\end{eqnarray} 
The corresponding dimension is $\delta_4=3$.

\begin{figure}
\centerline{\hskip.4in \epsffile{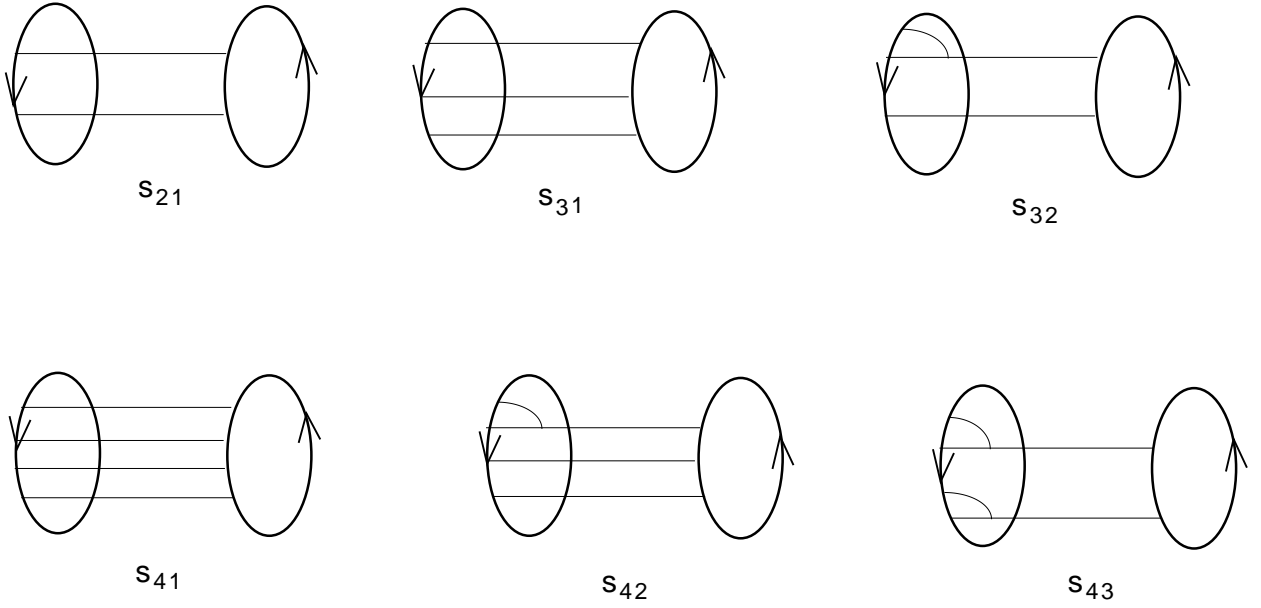}}
\caption{Basis of group factors up to order four.}
\label{closbasis.eps}
\end{figure}

Notice that, especially at order four, the dimensions have decreased 
significantly with respect to the case of open lines. The independent group 
factors that we have chosen are pictured in Fig.~\ref{closbasis.eps}. 
This choice plus the relations (\ref{primi}) lead to the following expressions
for the  geometrical factors:
\begin{eqnarray}
& & \gamma_{1}^{\,1} = \calc [\dga {1}^{\,1}], \nonumber \\ 
& & \gamma_{2}^{\,1} = { \calc [\dga {1}^{\,1}]^2  \over 2! }, \nonumber \\ 
& & \gamma_{3}^{\,1} = { \calc [\dga {1}^{\,1}]^3  \over 3! }, 
\;\;\;\;\; \gamma_{3}^{\,2} = \calc [\dga {1}^{\,1}]\calc [\dga {2}^{\,2}] 
+ 2 \calc [\dga {3}^{\,4}], \nonumber \\ 
& & \gamma_{4}^{\,1} = { \calc [\dga {1}^{\,1}]^4  \over 4! },
\;\;\;\;\; \gamma_{4}^{\,2} = { \calc [\dga {1}^{\,1}]^2  \over 2! }
\calc [\dga {2}^{\,2}] + \calc [\dga {1}^{\,1}] \calc [\dga {3}^{\,4}], 
\nonumber\\  
& & \gamma_{4}^{\,3} = { \calc [\dga {2}^{\,2}]^2  \over 2! } -
\calc [\dga {1}^{\,1}]\calc [\dga {3}^{\,3}] + \calc [\dga {4}^{\,7}] 
+ 2 \calc [\dga {4}^{\,8}] + 4 \calc [\dga {4}^{\,9}].  
\label{indgam}
\end{eqnarray} 
From these equations one can read the algebra of 
invariants for two-component links:
\begin{eqnarray}
& & \gamma_{2}^{\,1} = { (\gamma_{1}^{\,1})^2 \over 2!}, \nonumber \\ 
& & \gamma_{3}^{\,1} = { (\gamma_{1}^{\,1})^3 \over 3!}, \nonumber \\ 
& & \gamma_{4}^{\,1} = { (\gamma_{1}^{\,1})^4 \over 4!}, 
\;\;\;\;\; \gamma_{4}^{\,2} = { {\gamma_{1}^{\,1} \gamma_{3}^{\,2}} \over 2}, 
\label{alggam}
\end{eqnarray} 
so we have three primitive  invariants up to this
order: $\gamma_{1}^{\,1}$, 
$\gamma_{3}^{\,2}$ and  $\gamma_{4}^{\,3}$. Using this results we can finally 
write the power series expansion for 
$\langle \calz_{\lambda,\mu}(C_1,C_2,G)\rangle$ in (\ref{expintdos}) up to 
order four:
\begin{equation}
1 + x^2 \Big[{ (\gamma_{1}^{\,1})^2 \over 2!} s_{21} \Big] +
x^3 \Big[ { (\gamma_{1}^{\,1})^3 \over 3!} s_{31} +
 \gamma_{3}^{\,2} s_{32} \Big] +
x^4 \Big[ { (\gamma_{1}^{\,1})^4 \over 4!} s_{41}+
{ {\gamma_{1}^{\,1} \gamma_{3}^{\,2}} \over 2} s_{42} +
\gamma_{4}^{\,3} s_{43} \Big],
\label{laserie}
\end{equation}
where each of the quantities entering this expression are given in
(\ref{stwo}),  (\ref{sthree}), (\ref{sfour})  and (\ref{indgam}).

Notice that although there is no natural notion of primitiveness for the
group factors entering the power series expansion of 
$\langle \calz_{\lambda,\mu}(C_1,C_2,G)\rangle$ in (\ref{expintdos}), we
have obtained one for the geometrical factors from Eq.~(\ref{factores}).
This last equation is a consequence of the master equation
(\ref{mastereq}) and therefore of the  property of factorization of vacuum
expectation values in Chern-Simons gauge theory.

In the rest of this section we provide the explicit integral expressions of
the primitive invariants $\gamma_{1}^{\,1}$, 
$\gamma_{3}^{\,2}$ and  $\gamma_{4}^{\,3}$. There are two forms to face this
computation. The first one consists in writing down the relations for the
geometrical factors contained in (\ref{indgam}). The second form does not use
relations (\ref{indgam}) and one just writes down the general form of the
contribution in terms of Feynman diagrams for a product of Wilson loops at a
given order, organizing the expression so obtained in terms of independent
group factors. In this second approach one should obtain relations 
(\ref{alggam}) as a consequence. Indeed, we have verified that the adequate
parts of the  contributions factorize confirming  predictions (\ref{alggam})
from the factorization theorem. We will write down the expressions for the
primitives
$\gamma_{1}^{\,1}$, 
$\gamma_{3}^{\,2}$ and  $\gamma_{4}^{\,3}$ using the second approach.
Of course, for $\gamma_{1}^{\,1}$ both approaches  lead to the same expression.
For $\gamma_{3}^{\,2}$ and  $\gamma_{4}^{\,3}$, however, the expressions are
rather different. Its equivalence is one more prediction of the factorization
theorem. Though one could think that the first approach would lead to simpler
expressions, it turns out that this is not the case. The integral expressions
for $\dot \gamma_3^{\,\, 4}$, $\dot \gamma_4^{\,\, 7}$,
$\dot \gamma_4^{\,\, 8}$ and $\dot \gamma_4^{\,\, 9}$ are rather long as
compared to the ones obtained in the second approach. 

\begin{figure}
\centerline{\hskip.4in \epsffile{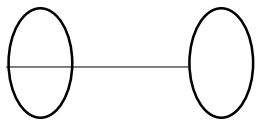}}
\caption{Diagrams contributing to $\gamma_{1}^{\,\,1}$.}
\label{gam11.eps}
\end{figure}

\begin{figure}
\centerline{\hskip.4in \epsffile{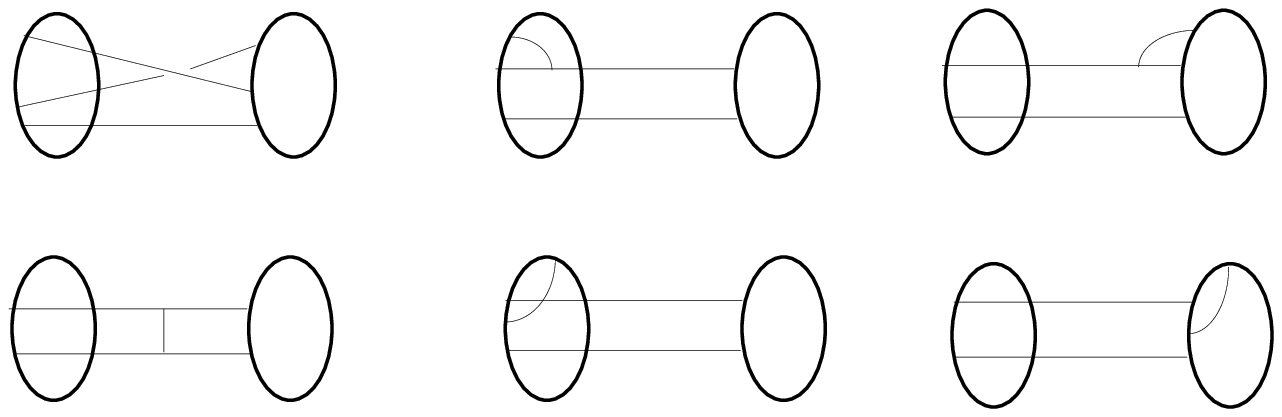}}
\caption{Diagrams contributing to $\gamma_{3}^{\,\,2}$.}
\label{gam32.eps}
\end{figure}

\begin{figure}
\centerline{\hskip.4in \epsffile{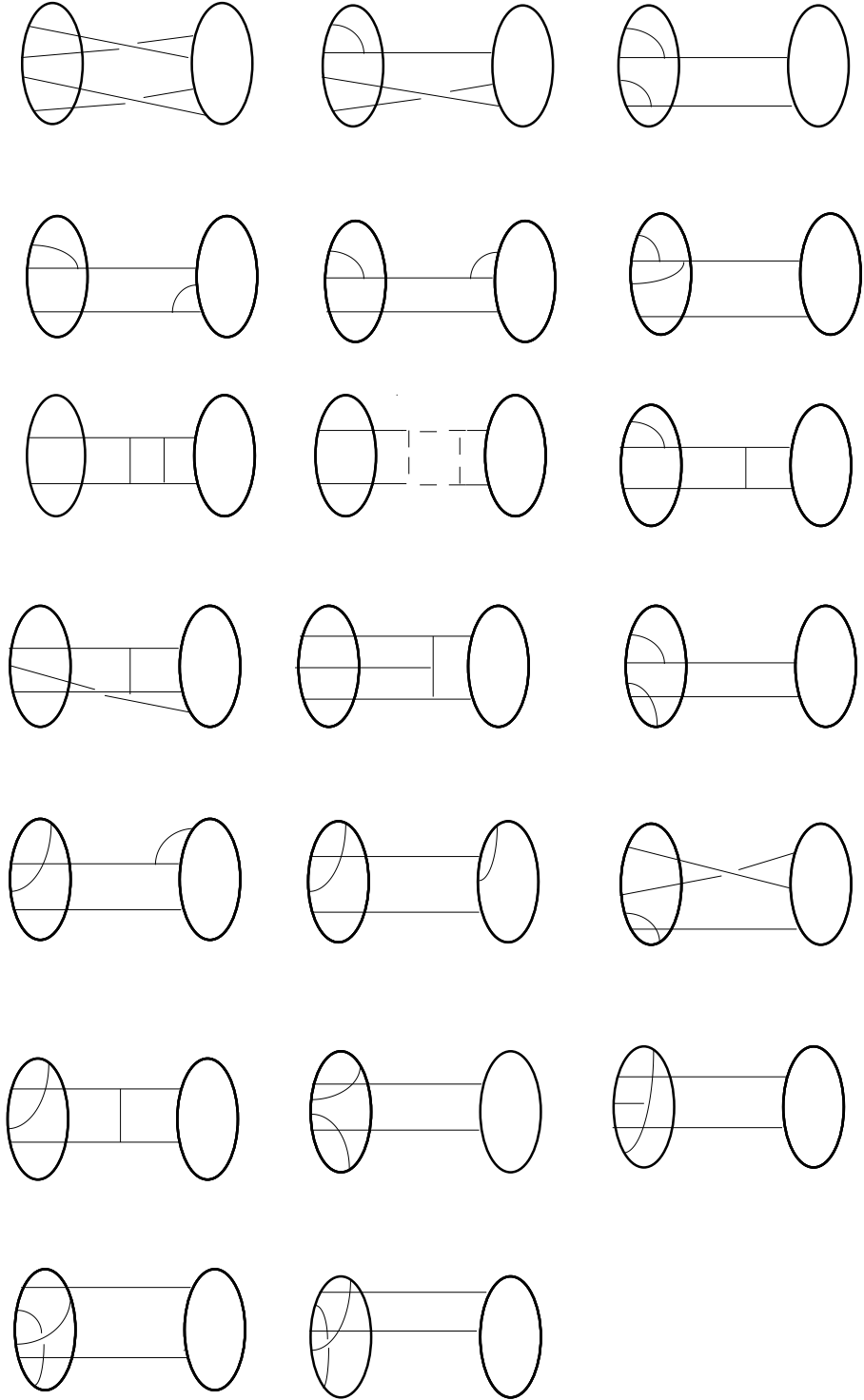}}
\caption{Diagrams contributing to $\gamma_{4}^{\,\,3}$.}
\label{gam43.eps}
\end{figure}

In order to obtain the  expresions for 
$\gamma_{1}^{\,1}$, 
$\gamma_{3}^{\,2}$ and  $\gamma_{4}^{\,3}$
one must take into account that each
term which contributes is  made up of the geometrical factor of a diagram 
whose group factor depends on the corresponding independent factor, multiplied 
by the constant that determines this dependence. The diagrams entering in
$\gamma_{1}^{\,1}$, 
$\gamma_{3}^{\,2}$ and  $\gamma_{4}^{\,3}$ are pictured in 
Figs.~\ref{gam11.eps}, \ref{gam32.eps} and \ref{gam43.eps} 
respectively. All integration paths are taken anticlockwise. Before writting 
down these expresions, we will introduce some notation that will simplify them
considerably. (The number of {\it lines} needed to write the integral 
$\ga {3}^{\,\,2}$ is big, but still reasonable. But for $\ga {4}^{\,\,3}$ the 
number of {\it pages} will certainly be unreasonable). We shall write the 
multiple integral over the first Wilson loop as, 
\beq
 \oint dx_n^{\mu_n} 
 \int^{x_n} dx_{n-1}^{\mu_{n-1}} \cdots  \int^{x_2} dx_1^{\mu_1}
 \equiv  \oint_{1<2< \cdots <n} dx ,
\label{intx}
\eeq 
while the one over the second loop,
\beq
 \oint dy_m^{\nu_m} 
 \int^{y_m} dy_{m-1}^{\nu_{n-1}} \cdots  \int^{y_2} dy_1^{\nu_1}
 \equiv    \oint_{1<2< \cdots <m} dy.
\label{inty}
\eeq 
In these equations $n$ and $m$ label the number of points over the first and
second Wilson loops, respectively. Also, the variable $x$  will always
correspond to points attached to the first loop, and the variable $y$ to the
second. Internal vertices will be labelled by variables $\omega_i$. One finds
four types of propagators,
\begin{eqnarray}
& & p(x_i,y_j) = \Delta_{\mu_i \nu_j} (x_i - y_j),
\;\;\;\;\; p(\omega_i,\omega_j) 
= \Delta_{\theta_i \theta_j} (\omega_i - \omega_j),
\nonumber \\ 
& & p(x_i,\omega_j) = \Delta_{\mu_i \theta_j} (x_i -
\omega_j),
\;\;\;\;\; p(\omega_i,y_j) = \Delta_{\theta_i \nu_j} (\omega_i -
y_j), 
\label{propa}
\end{eqnarray}
depending if they connect points on the Wilson lines, two internal vertices,
or one point on a Wilson line and the other on an internal vertex.
In (\ref{propa}) $\Delta_{\mu \nu} (x - y) $ is 
the quantity defined in (\ref{elpropa}).

From the Feynman rules, and taking into account that one has to
extract a factor $x^i$ at order $i$ where $x=ig^2/2$, one finds the following
set of effective rules for the computation of the geometrical factors
associated to the diagrams shown in Figs.~\ref{gam11.eps}, \ref{gam32.eps}
and \ref{gam43.eps}:
\vskip.25cm
\noindent - One $p(,)$ as in (\ref{propa}) for each internal line,
whose variables are fixed by its end points, with a factor ${1\over 4}$ for
each.

\noindent - Two path ordered integrals, one over the $n$ points on the
first loop, and the other over the $m$ on the second.

\noindent - A factor $\epsilon^{\alpha_i\beta_i \gamma_i} = \epsilon_i$
and a three-dimensional integral $\int {\rm d}^3 \omega_i$, for each
three-vertex, with the lines in the vertex ordered counterclokwise. The
 $p(,)$ attached to the vertex will be written in this order so that
one can keep track of it.

\noindent - A factor $2^i$ , where $i$ is the order in perturbation theory.

\noindent - Finally, there might be a multiplicative factor coming
from the relation between the group factor of a given diagram and the one
chosen as independent.

Notice that in these rules there is not a factor of the form $i$ to some power
as one could naively expect form the Feynman rules. The reason for this is that
the resulting power of $i$ is always absorbed in the parameter $x$ at each order
with no sign left.

\vskip 0.25cm
The integral expresion for $\gamma_{1}^{\,1}$ is easily obtained applying these
rules to the diagram shown in Fig.~\ref{gam11.eps}:
\beq
\gamma_{1}^{\,1} ={1 \over 2} \oint {\rm d}x \oint {\rm d}y \; p(x,y).
\label{ga11}
\eeq 
This expression is twice the integral defining the linking number of the two
components of the link under consideration. Therefore, our first numerical
link invariant turns out to be one of the classical link invariants. At higher
orders, however, the numerical invariants that we present are new.

For the only primitive invariant at order three (recall that
there are not primitive invariants at order two), $\gamma_{3}^{\,2}$, the
corresponding diagrams are depicted in Fig.~\ref{gam32.eps}. In writing 
their corresponding integrals it is important to take into account the 
following two observations. First, notice that if the whole diagram
but the two Wilson loops is not symmetric under a reflection around its medium
vertical axis, the diagram obtained reflecting the graph also contributes.
Second, each diagram represents a whole class of contributions: all the
distinct  contributions which can be obtained permuting the order in which the
legs  are attached to the Wilson loops. One finds:
\begin{eqnarray}
\gamma_{3}^{\,2}  =  {1 \over 8} & &\oint_{1<2<3} dx  \oint_{1<2<3} dy \bigg[
                 p(x_1,y_2) p(x_2,y_1) p(x_3,y_3)  + \nonumber \\
             & &  p(x_1,y_1)p(x_2,y_3)p(x_3,y_2) +
p(x_1,y_3)p(x_2,y_2)p(x_3,y_1) \bigg]  \nonumber \\
  + {1 \over 32} & &\oint_{1<2<3}dx \oint_{1<2} dy \int d^3\omega \;
\epsilon \bigg[
    p(x_2,\omega)p(x_1,\omega)p(x_3,y_2)p(\omega,y_1) + \nonumber \\
   & &  p(x_1,y_1)p(x_2,\omega)p(x_3,\omega)p(\omega,y_2) +
    p(x_1,\omega)p(x_2,\omega)p(x_3,y_1)p(\omega,y_2) + \nonumber \\
   & &  p(x_1,y_2)p(x_2,\omega)p(x_3,\omega)p(\omega,y_1) +
    p(x_3,\omega)p(x_1,\omega)p(x_2,y_2)p(\omega,y_1) + \nonumber \\
  & &  p(x_3,\omega)p(x_1,\omega)p(x_2,y_1)p(\omega,y_2) \bigg] \nonumber \\
-{1 \over 32}& & \oint_{1<2} dx \oint_{1<2<3} dy \int d^3\omega \;
\epsilon \bigg[
       p(x_1,\omega)p(\omega,y_2)p(\omega,y_1)p(x_2,y_3) + \nonumber \\
     & &  p(x_1,y_1)p(x_2,\omega)p(\omega,y_3)p(\omega,y_2) +
       p(x_1,y_3)p(x_2,\omega)p(\omega,y_2)p(\omega,y_1) + \nonumber \\
     & &  p(x_1,\omega)p(x_2,y_1)p(\omega,y_3)p(\omega,y_2) +
       p(x_1,\omega)p(x_2,y_2)p(\omega,y_1)p(\omega,y_3) + \nonumber \\
  & &  p(x_1,y_2)p(x_2,\omega)p(\omega,y_1)p(\omega,y_3) \bigg] \nonumber \\
+{1 \over 64} & & \oint_{1<2} dx \oint_{1<2} dy \int d^3\omega_1 \int
d^3\omega_2 \; \epsilon_1 \epsilon_2 \nonumber \\
  & & \bigg[
p(x_1,\omega_1)p(\omega_1,\omega_2)p(x_2,\omega_2)p(\omega_2,y_2)
  p(\omega_1,y_1) + \nonumber \\  & & 
p(x_1,\omega_2)p(x_2,\omega_1)p(\omega_2,y_2)p(\omega_1,\omega_2)
p(\omega_1,y_1) \bigg]  \nonumber \\
 -{1 \over 8} & & \oint_{1<2<3<4}dx \oint_{1<2} dy
\bigg[ p(x_1,x_3)p(x_2,y_1)p(x_4,y_2) + \nonumber \\
 & &  p(x_1,y_1)p(x_2,x_4)p(x_3,y_2) + p(x_1,x_3)p(x_2,y_2)p(x_4,y_1) +
\nonumber \\
  & & p(x_1,y_2)p(x_2,x_4)p(x_3,y_1) \bigg]  \cr  -{1 \over 8}  & &
\oint_{1<2}dx \oint_{1<2<3<4} dy \bigg[ p(x_1,y_2)p(x_2,y_4)p(y_1,y_3)  +
\nonumber \\
 & & p(x_1,y_1)p(x_2,y_3)p(y_2,y_4) + p(x_1,y_4)p(x_2,y_2)p(y_1,y_3) +
\nonumber \\
  & & p(x_1,y_3)p(x_2,y_1)p(y_2,y_4) \bigg].  
\label{gam32}
\end{eqnarray}
In (\ref{gam32}) we have written the integrands in such a way that the relative
minus signs between the different terms in a given integral, which may arise
from the group factor dependence of some of the permutations, are reabsorbed by
the order in which propagators are written. The overall sign before each
integral is taken to be the one associated to the choice of permutation drawn
in the figures.

Before writting the integral expresion for $\gamma_{4}^{\,3}$ we
are going to introduce a new notation which notably simplifies our formulae.
We will make the substitution:
\begin{eqnarray} 
& & p(x_i,y_j) \too \theta_{i,\bar j}\; , \nonumber \\ & &
p(\omega_i,y_j) \too \theta_{i,\bar j} ,
\label{tita}
\end{eqnarray} 
in such a way that indices with a bar above them label the variables $y$ ,
while the others label the variables $x$ or $\omega$. There is no confusion
between them as the indices from the $\omega$ variables will appear
repeated three times. Of course, the $\theta$'s will be written preserving the
order of the $p(,)$'s, which is that of the propagators entering in each 
three-vertex. Given a topology, for example any of the diagrams pictured in 
Fig.~\ref{gam32.eps}, our integral has a term for each of the permutations 
whose group factor gives a contribution to the invariant. These permutations 
are realized in a given order of the variables $\{ x_1$,\dots,$x_n$; $y_1$,
\dots,$y_m$; $\omega_1$,\dots,$\omega_t \}$ in the integrand. One can always 
choose one of them as a reference, and make a change of variables in the 
others, applying the inverse of the given permutation in each case. We will 
end up with a sum of integrals with the same integrand but different domains 
of integration:
\begin{eqnarray}
 & & \oint_{1<2< \cdots <n} dx  \oint_{1<2< \cdots <m} dy 
\int d\omega_1 \cdots  d\omega_t \;
\sum\limits_{\sigma}f \big(\sigma(x_1 ... \omega_t) \big)
\nonumber \\  & & = \sum\limits_{\sigma^{-1}}
\oint_{\sigma^{-1}(1< \cdots <n)} dx 
\oint_{\sigma^{-1}(1<
\cdots < m)} dy 
\int d\omega_{\sigma^{-1}(1)} \cdots  d\omega_{\sigma^{-1}(t)} \; f
\big(x_1,..., \omega_t \big)
\label{sigmas}
\end{eqnarray} 
We will write this integral in a more compact form,
defining the domain of integration in the following way
\beq
\int_{C_{n,m}} f = {1 \over N_d}
\sum\limits_{\sigma^{-1}}
\oint_{\sigma^{-1}(1< \cdots <n)} dx 
\oint_{\sigma^{-1}(1<
\cdots <m)} dy 
\int d\omega_{\sigma^{-1}(1)} \cdots \int d\omega_{\sigma^{-1}(t)} f
\label{redef}
\eeq
where $N_d$ is the number of different domains. So  
$C_{n,m}$ represents a  sum of integrals of dimensions $n+m+3t$, where the
value of $t$ is read from the number of repeated indices in the integrand,
which is a product of the $\theta$'s  defined in (\ref{tita}). Using
this notation we can rewrite the invariant $\gamma_{3}^{\,2}$   in
(\ref{gam32}) as:
\begin{eqnarray}
\gamma_{3}^{\,2} = {3 \over 8}& & \int_{C_{3,3}}
\theta_{1,\bar2}\theta_{2,\bar1}
       \theta_{3,\bar 3} 
       +{3 \over 16} \int_{C_{3,2}}\theta_{2,4}\theta_{1,4}\theta_{3,\bar 2}
       \theta_{4,\bar 1} 
       - {3 \over 16} \int_{C_{2,3}} \theta_{1,4}
       \theta_{4,\bar 2}\theta_{4,\bar 1}\theta_{2,\bar 3} \nonumber \\
     +{1 \over 32} & & \int_{C_{2,2}} \theta_{1,3}\theta_{3,4}\theta_{2,4}
     \theta_{4,\bar 2}\theta_{3,\bar 1} 
     - {1 \over 2} \int_{C_{4,2}} \theta_{1,3}\theta_{2,\bar 1}
     \theta_{4,\bar 2} - {1 \over 2} \int_{C_{2,4}} \theta_{1,\bar 2}
      \theta_{2,\bar 4}\theta_{\bar 1,\bar 3}
 \label{gambis32}
 \end{eqnarray}

Let us now write down the integral expresion for $\gamma_{4}^{\,3}$. Diagrams
contributing to it, up to permutations, are pictured in 
Fig.~\ref{gam43.eps}. For a non symmetric Feynman graph, the reflected 
diagram also contributes, although we have not drawn it. Notice that at this 
order we encounter the first diagram coming from the ghost sector, which is 
the one with dashed lines in the figure. The last three diagrams are considered
different topologies, because of the different relative position between the 
interaction and the loop subdiagrams.

As for the ghost diagram, we follow similar conventions:
$\hat \theta$ stands for the ghost propagator. The three-vertex between the
ghosts and gauge fields introduces, besides the three dimensional integration, a
derivative in one of the variables in the ghost propagator. Recall that this
vertex is not antisymmetric, so the order in which propagators  are written is
not important.

\beq
\hat \theta_{i,j} = \partial_i D(x_i - y_j)
\label{ghost}
\eeq
where $D(x_i - y_j)$ is the geometrical part of the ghost propagator written in
(\ref{kernels}). The order of the subindices in $\hat \theta_{i,j}$ keeps track
of the variable on which the derivative acts.

The kind of multiplicative factors entering in this term are
of the same nature as before, \ie\ : $2^i$ for order $i$ perturbation theory,
$1/4$ for each propagator, and the factors coming from the group factor
dependence and the number of different domains (or diagrams contributing to that
topology).  There is also an extra factor 2 coming from the fact that the two
orientations of the ghost loop must be taken into account. Factors of $i$ in
diagrams involving ghosts are also reabsorbed in the perturbation parameter
with no sign left.

Using the rules described above our expresion for the  invariant
$\gamma_{4}^{\,3}$ is the following:
\begin{eqnarray}
\gamma_{4}^{\,3} = {1 \over 4} & & \int_{C_{4,4}} \theta_{1,\bar2}
      \theta_{2,\bar1}\theta_{3,\bar 4}\theta_{4,\bar 3}
     +  {7 \over 64}  \int_{C_{4,3}} \theta_{1,\bar2}
      \theta_{2,\bar1}\theta_{3,5}\theta_{5,\bar 3}\theta_{4,5}
     +  {7 \over 64}  \int_{C_{3,4}} \theta_{1,\bar2}
      \theta_{2,\bar1}\theta_{\bar3,5}\theta_{\bar4,5}\theta_{5,3}
      \nonumber \\
     + {1 \over 64} & & \int_{C_{4,2}} \theta_{1,5} \theta_{5,\bar1}
      \theta_{2,5} \theta_{3,6} \theta_{6,\bar2} \theta_{4,6}
     + {1 \over 64} \int_{C_{2,4}} \theta_{5,\bar1} \theta_{5,\bar2}
      \theta_{1,5} \theta_{6,\bar3} \theta_{6,\bar4} \theta_{1,6} 
     - {9 \over 256} \int_{C_{3,3}} \theta_{1,4}\theta_{4,\bar1}
     \theta_{4,\bar2}\theta_{2,5}\theta_{5,\bar3}\theta_{3,5}
     \nonumber \\
     - { 9 \over 256} & & \int_{C_{3,3}}\theta_{1,\bar1}\theta_{2,5}
     \theta_{5,6}\theta_{5,3}\theta_{6,\bar2}\theta_{6,\bar1}
     +{3 \over 64} \int_{C_{4,2}} \theta_{1,\bar1}\theta_{2,5}
     \theta_{5,\bar2}\theta_{5,6}\theta_{6,4}\theta_{6,3}
      +{3 \over 64} \int_{C_{2,4}} \theta_{1,\bar1}\theta_{2,5}
     \theta_{5,\bar2}\theta_{5,6}\theta_{6,\bar3}\theta_{6,\bar4}
     \nonumber \\
     + {1 \over 512} & & \int_{C_{2,2}}\theta_{1,3}\theta_{3,4}
     \theta_{3,6}\theta_{6,5}\theta_{6,2}\theta_{4,\bar1}\theta_{4,5}
     \theta_{5,\bar2}
     + { 1 \over 64} \int_{C_{3,2}} \theta_{1,4} \theta_{4,\bar1}
     \theta_{4,5}\theta_{5,\bar2}\theta_{5,6}\theta_{6,3}\theta_{6,2}
     \nonumber \\
     + { 1 \over 64} & & \int_{C_{2,3}} \theta_{1,4}\theta_{4,\bar1} 
     \theta_{4,5}\theta_{5,6}\theta_{5,2}\theta_{6,\bar2}\theta_{6,\bar3}
     + {1 \over 32}  \int_{C_{3,3}} \theta_{1,4}\theta_{4,\bar2}
     \theta_{4,5}\theta_{2,\bar1}\theta_{5,\bar3}\theta_{5,3}
     \nonumber \\
     + {3 \over 512} & & \int_{C_{3,2}} \theta_{1,4}\theta_{4,\bar1}
     \theta_{4,5}\theta_{5,6}\theta_{5,2}\theta_{6,\bar2}\theta_{6,3}
     - {3 \over 512} \int_{C_{2,3}} \theta_{1,4}\theta_{4,\bar1}
     \theta_{4,5}\theta_{5,\bar2}\theta_{5,6}\theta_{6,\bar3}\theta_{6,2}
     \nonumber \\
     - {15 \over 32} & & \int_{C_{5,2}} \theta_{1,3}\theta_{2,\bar1}
     \theta_{4,6}\theta_{6,\bar2}\theta_{6,5}
     - {15 \over 32}  \int_{C_{2,5}} \theta_{\bar1,\bar3}\theta_{1,\bar2}
     \theta_{6,\bar4}\theta_{6,\bar5}\theta_{2,6}
     + { 3 \over 16} \int_{C_{4,3}}\theta_{1,\bar1}\theta_{2,4}
     \theta_{3,5}\theta_{5,\bar2}\theta_{5,\bar3}
     \nonumber \\
     + { 3 \over 16} & & \int_{C_{3,4}}\theta_{1,\bar1}\theta_{\bar2,\bar4}
     \theta_{5,\bar3}\theta_{3,5}\theta_{2,5}
     + { 1 \over 2} \int_{C_{4,4}}\theta_{1,\bar1}\theta_{2,4}
     \theta_{3,\bar3}\theta_{\bar2,\bar4}
     - { 15 \over 16} \int_{C_{5,3}}\theta_{1,3}\theta_{2,\bar1}
     \theta_{4,\bar3}\theta_{5,\bar2}
     \nonumber \\
     - { 15 \over 16} & & \int_{C_{3,5}}\theta_{\bar1,\bar3}
     \theta_{1,\bar2}\theta_{2,\bar5}\theta_{3,\bar4}
     - {1 \over 32} \int_{C_{4,2}}\theta_{1,5}\theta_{5,\bar1}
     \theta_{5,6}\theta_{6,\bar2}\theta_{6,3}\theta_{2,4}
     \nonumber \\
     - {1 \over 32} & & \int_{C_{2,4}}\theta_{1,5}\theta_{5,\bar1}
     \theta_{5,6}\theta_{6,\bar3}\theta_{6,2}\theta_{\bar2,\bar4}
     + {3 \over 8} \int_{C_{6,2}} \theta_{1,3}\theta_{2,\bar1}
     \theta_{4,6}\theta_{5,\bar2}
     + {3 \over 8} \int_{C_{2,6}} \theta_{\bar1,\bar3}\theta_{1,\bar2}
     \theta_{\bar4,\bar6}\theta_{2,\bar5}
     \nonumber \\
     -{5 \over 32} & & \int_{C_{5,2}} \theta_{2,\bar1}\theta_{4,\bar2}
     \theta_{1,6}\theta_{3,6}\theta_{5,6}
     -{5 \over 32}  \int_{C_{2,5}} \theta_{1,\bar2}\theta_{2,\bar4}
     \theta_{6,\bar1}\theta_{6,\bar5}\theta_{6,\bar3}
     + {3 \over 4} \int_{C_{6,2}} \theta_{1,4}\theta_{2,\bar1}
     \theta_{3,6}\theta_{5,\bar2}
     \nonumber \\
     + {3 \over 4} & & \int_{C_{2,6}} \theta_{\bar1,\bar4}\theta_{1,\bar2}
     \theta_{\bar3,\bar6}\theta_{2,\bar5}
     + {1 \over 2} \int_{C_{6,2}} \theta_{1,4}\theta_{2,6}
     \theta_{3,\bar1}\theta_{5,\bar2}
     + {1 \over 2}  \int_{C_{2,6}} \theta_{\bar1,\bar4}
     \theta_{\bar2,\bar6}\theta_{1,\bar3}\theta_{2,\bar5}
      \nonumber \\
     + {1 \over 256} & & \int_{C_{2,2}}\theta_{1,3}\theta_{2,4}\theta_{\bar1,6}
    \theta_{\bar2,5}\hat\theta_{3,4}\hat\theta_{4,5}\hat\theta_{5,6}
     \hat\theta_{6,3}.
 \label{gam43}
 \end{eqnarray}
 where the last term stands for the ghost diagram contribution.
     
\newpage
\setcounter{equation}{0}
\section{Numerical link invariants for links up to six crossings}

In this section we present the results of a numerical computation of the 
Vassiliev invariants for links up to six crossings up to order four. Although 
we know integral expressions for the $\gamma_{i}^{\,j}$, we will not
evaluate them, as we have a shorter way to proceed. The computation will be
carried out  using information of the l.h.s. of (\ref{closed}) coming from the
polynomial  invariants for links. Given a link, we may use the polynomials
defined for  different Lie groups and representations, and compare them with 
the r.h.s. of  (\ref{closed}). As all the group dependence is encoded in the
$R_{ij}$, and the $\gamma_{i}^{\,j}$ only depend on the link under
consideration, we end with a set of linear equations for them. 
As in \cite{numbers} we could consider
the following cases: $SO(N)$ in its  fundamental representation (Kauffman
polynomial
\cite{kauf}), $SU(N)$ in its  fundamental representation (HOMFLY polynomial
\cite{homfly}), $SU(2)_j$ in an  arbitrary spin $j$ representation (Jones and
Akutsu-Wadati polynomials  \cite{jones,aw,kaul,pol}), and $SU(N) \times
SO(N)$ also in the fundamental representation.  All these  invariants are known
and can be collected from the  literature (for example, \cite{millet} and
\cite{kaul}). At the order in perturbation theory considered in this work,
however, it is enough to consider the HOMFLY and the Kauffman polynomials for
the links under study. They are listed in Appendix C.

The structure of the computation to be carried out is as follows. Once the
polynomial invariant corresponding to the l.h.s. of (\ref{closed}) is
collected, one replaces the variable $q$ by $e^x$ and expands in powers of 
$x$. For the case considered here we only need the expansion up to order 
four. The coefficients of $x^i$ are either polynomials in $N$ or polynomials 
in $j$. On the other hand, in the r.h.s. of (\ref{closed}), the group 
factors are the ones given diagrammaticaly in Fig.~\ref{closbasis.eps}; their
explicit expressions are written in Appendix A. Again, these group factors are 
polynomials in $N$ or $j$. Comparing both sides of (\ref{closed}) leads to a 
series of linear equations for the geometrical factors
$\gamma_{i}^{\,j}$ that  determine uniquely all the  $\gamma_{i}^{\,j}$
up to order four. These values  are listed in the following table. One can 
check that the algebraic relations  in (\ref{alggam}) hold, and that the 
$\gamma_{i}^{\,j}$ are all rational numbers. The notation used to label links 
is the same one as in  \cite{millet}. The big number  represents the number of
crossings, the superscript the number of link  components, and the subscript
refers just to a standard ordering of links with  a given number of crossings.
The sign stands  for the relative orientation  between the two components: it 
is $+$ when the  linking number is taken  positive, and $-$ otherwise. If 
there is no sign, one of the  links is obtained from the other by reversing 
the orientation of space. While the invariants of even order remain unchanged 
after reversing the orientation of space, the ones of order odd change sign.

\vskip 0.5cm
\begin{center}
\begin{tabular}{|c|c|c|c|c|c|c|c|c|c|} \hline
 Link & $\gamma_{1}^{\,\,1}$ & $\gamma_{2}^{\,\,1}$ & $\gamma_{3}^{\,\,1}$ 
& $\gamma_{3}^{\,\,2}$
& $\gamma_{4}^{\,\,1}$ & $\gamma_{4}^{\,\,2}$ & $\gamma_{4}^{\,\,3}$  \\ \hline
$2^2_1$ & $2$ & $2$ & $4/3$ & $2/3$ & $2/3$
& $2/3$ & $0$ \\ \hline
$4^2_{1+}$ & $4$ & $8$ & $32/3$ & $28/3$ & $32/3$
& $56/3$ & $8$ \\ \hline
$4^2_{1-}$ & $-4$ & $8$ & $-32/3$ & $-4/3$ & $32/3$
& $8/3$ & $0$ \\ \hline
$5^2_{1}$ & $0$ & $0$ & $0$ & $-8$ & $0$
& $0$ & $-8$ \\ \hline
$6^2_{1+}$ & $6$ & $18$ & $36$ & $34$ & $54$
& $102$ & $48$ \\ \hline
$6^2_{1-}$ & $-6$ & $18$ & $-36$ & $-2$ & $54$
& $6$ & $0$ \\ \hline
$6^2_2$ & $6$ & $18$ & $-36$ & $-18$ & $54$
& $54$ & $16$ \\ \hline
$6^2_{3+}$ & $4$ & $8$ & $32/3$ & $52/3$ & $32/3$
& $104/3$ & $32$ \\ \hline
$6^2_{3-}$ & $-4$ & $8$ & $-32/3$ & $20/3$ & $32/3$
& $-40/3$ & $8$ \\ \hline
\end{tabular}
\end{center}
\centerline{Table}
\vskip 1cm

Recall that only $\gamma_1^{\,\,1}$, $\gamma_3^{\,\,2}$
and $\gamma_4^{\,\,3}$ are primitve invariants. The results
presented in the table are consistent with our interpretation
of the invariant $\gamma_1^{\,\,1}$ as a twice the linking number
of the two components. It is not clear, however, if the other
two, $\gamma_3^{\,\,2}$ and $\gamma_4^{\,\,3}$ are related to some
known numerical invariants. 

\newpage
\section{Conclusions}

In this paper we have presented a generalization of the approach
introduced in \cite{numbers} for knots to the case of two-component links.
There is a very fundamental difference between both approaches since the 
simple algebraic structure for group factors in the case of knots is not
present for links. There is, however, a rather similar algebraic structure
if one considers open links. This fact leads us to introduce an opening and
closing operation and to deal first with  the analysis of the algebraic
structure for open links. After showing the commutative nature of the
tensor product involved in that algebraic structure we are able to 
prove a factorization theorem similar to the one first introduced in 
\cite{factor}. This theorem allows, on the one hand, to factorize the 
dependence on numerical knot invariants corresponding to each of the 
components, and therefore to isolate the pure link contributions, on the other 
hand, to define a
notion of primitiveness for these pure link invariants. After closing, the
relations  derived from the factorization theorem play a fundamental role to
obtain relations among numerical invariants for links, and a notion of
primitiveness is also defined for these invariants. The number $\hat\delta_i$
of primitive numerical link invariants  at order $i$ are 1, 0, 1 and 1, for 
$i=1,\dots,4$ respectively. The first one can be
identified with twice the linking number of the two components of the link. 

Using the theorems in \cite{birlin,birman} we have been able to argue that
the geometrical factors $\gamma_i^{\,\,j}$ are Vassiliev invariants or
numerical invariants of finite type. Our work provides explicit integral
expressions for some of the simplest cases. We have not been able to identify
the invariants $\gamma_3^{\,\,2}$ and $\gamma_4^{\,\,3}$ with known ones.

We have used the methods first introduced in \cite{numbers} to compute 
invariants up to order four for some simple links. It turns out that
up to this order it is enough to use the information derived from the HOMFLY
and the Kauffman polynomials. We have obtained a set of rational numbers which
are consistent with (\ref{alggam}), which is a direct consequence of the
factorization theorem. It is not clear from our results that a
natural normalization which makes all the  primitive invariants integer-valued
(as in the case of knot
invariants \cite{numbers,torus}) exists.

This work opens a variety of investigations. First, one should try to compute
higher-order invariants. The cases studied in this work are the simplest
ones and should be considered as examples to implement our program for
numerical invariants. Second, the approach should be generalizaed to
$n$-component links. In particular, one should study how the factorization
theorem generalizes in that case and how many classes of invariants can be
defined. Third, one should analyze the properties of the $\dot
\gamma_{i}^{\,\,j}$. These quantities are not ambient isotopy invariants since
they are gauge dependent but they might be interesting in some gauges if one is
able to show that they preserve some form of isotopy. This seems to be the case
as shown in \cite{wetering} and the resulting invariants might be related to
string link invariants \cite{barstring}. Finally, the invariants presented in
this paper should be also regarded from the approach proposed in \cite{bot}. We
plan to report on these and some other issues related to Vassiliev invariants in
future work.

\newpage

\vskip 1cm                                               
{\Large{\bf APPENDIX A}}                                 
\vskip .5cm                                              
\renewcommand{\theequation}{\rm{A}.\arabic{equation}}    
\setcounter{equation}{0}                                 

In this appendix we present a summary of our group-theoretical conventions.
 We choose the generators of the Lie algebra $A$ to be antihermitian such that
\beq
[T^a,T^b] = - f^{ab}_{\,\,\,\,\, c}T^c,
\label{auno}
\eeq
 where $f^{ab}_c$ are the structure constants.  These satisfy the
Jacobi identity,
\beq
f^{ab}_{\,\,\,\,\,e}f^{ec}_{\,\,\,\,\,d}+f^{cb}_{\,\,\,\,\,e}
f^{ae}_{\,\,\,\,\,d}
+f^{ac}_{\,\,\,\,\,e}f^{be}_{\,\,\,\,\,d}=0
\label{jacobi}
\eeq
The generators are normalized in such a way that for the fundamental
representation,
\beq
\tr(T_aT_b)=-{1\over 2} \delta_{ab},
\label{normagene}
\eeq
 where $\delta_{ab}$ is the Kronecker delta. This can always be done
for  compact semisimple Lie algebras which is the case considered in
this paper. 

The generators $T^a$ in the adjoint representation coincide with the
structure constants,
\beq
\big(T^a\big)_c{}^b=f^{ab}_{\,\,\,\,\,c}\qquad {\hbox{\rm (adjoint
representation).}}
\label{adjunta}
\eeq
 The quadratic Casimir in the adjoint representation, $C_A$, is 
defined as
\beq
f^{ad}_{\,\,\,\,\,c} f^{bc}_{\,\,\,\,\,d}=C_A \delta^{ab}.
\label{Killingone}
\eeq
 The value of $C_A$ for the groups $SU(N)$ and $SO(N)$ is $-N$ and
$-{1\over 2}(N-2)$ respectively. The Killing metric is chosen to be the
identity matrix and therefore one can lower and raise group indices
freely.  For the case under consideration 
$f_{abc}$ is totally antisymmetric.

The convention chosen in (\ref{auno}) seems unusual but it is the 
most convenient
when  the Wilson line is defined as in (\ref{wilsonloop}). If we  had chosen
$if^{abc}$ instead of $-f^{abc}$, the exponential of the Wilson line
would have had $ig$ instead of $g$. Our convention also introduces a
$-1$ in the gauge vertex (\ref{kernels})

Our aim is to calculate the independent group structures appearing
in the perturbation series expansion of a two-component link (\ref{closed}), for
some given representations of the Lie algebra. For the factorized knot part, they
correspond to the Casimirs of that algebra, and where already worked out in
\cite{numbers}. Here we will restrict ourselves to the pure link factors. They
can be thought as generalized Casimirs, living in the tensor product of two
representations of the envelopping algebra, as they general form is a product 
of two traces over invariant tensors:
\beq
s = \tr_{R_1} (f^{abc} T_d^{(R_1)} T_e^{(R_1)} \cdots )
\tr_{R_2} (f^{qrs} T_u^{(R_2)} T_v^{(R_2)} \cdots ) 
\label{casgen}
\eeq
where all the indices are repeated and summed over. The evaluation procedure is
similar to the  case for knots \cite{vita,nordita,numbers}. First, we get rid
of the structure constants using the
conmutation relations, so we end with a sum of terms with the form 
(\ref{casgen}) but made only of products of generators. Secondly, we need
some expression for the tensor product of generators with the same index in
arbitrary representations.
\beq
\bigg( T_a^{(R_1)} \bigg)_i^{\;\,j} \otimes \bigg( T_a^{(R_2)} 
\bigg)_k^{\;\,l}
\label{proyector}
\eeq
These group-theoretical objects are called projection operators. They
are explicitly known for every classical Lie group except
$E_8$ \cite{vita,nordita}  when both, $R_1$ and $R_2$, stand for the fundamental
representation. In the case of
$SU(N)$ the projection operator is: 
\beq
\Big(T_a\Big)_i^{\,\,\,j}\Big(T_a\Big)_k^{\,\,\,l}=-{1\over
2}\Big(\delta_i^{\,\,\,l}\delta_k^{\,\,\,j}-{1\over
N}\delta_i^{\,\,\,j}\delta_k^{\,\,\,l}
\Big),
\label{rulesun}
\eeq
while for $SO(N)$,
\beq
\Big(T_a\Big)_i^{\,\,\,j}\Big(T_a\Big)_k^{\,\,\,l}=-{1\over
4}\Big(\delta_k^{\,\,\,j}\delta_i^{\,\,\,l}-\delta^{jl}\delta_{ik}\Big).
\label{ruleson}
\eeq
Similar identities can be read from \cite{nordita} for other
groups. This solves the problem of calculating the group factors in the
fundamental representation of these groups.

Higher representations can be introduced as properly symmetrized
products of fundamental representations. These products span the
representation ring of any compact Lie group. Extensions of (\ref{rulesun})
and (\ref{ruleson}) can be found, which would enable us to evaluate the analog
of (\ref{proyector}) in these more involved cases.

Using these rules we have computed all the independent group factors $s_{ij}$
up to order four for the simple Lie algebras which have been used in this
paper. They correspond to the fundamental representations of
$SU(N)$ and  $SO(N)$. The general form of the independent group 
factors pictured
in  Fig.~\ref{closbasis.eps} are:
\begin{eqnarray}
s_{21} &=& {1 \over {d_1 d_2}} \tr_{R_1} \big(  T_a T_b \big)
\tr_{R_2} \big( T_aT_b \big) \nonumber \\
s_{31} &=& {1 \over {d_1 d_2}} \tr_{R_1} \big( T_b T_c T_a\big) 
\tr_{R_2} \big( T_aT_bT_c \big) \nonumber \\
s_{32} &=& -{1 \over {d_1 d_2}} f^{dea} \tr_{R_1} \big( T_b T_e T_d\big) 
\tr_{R_2} \big( T_aT_b \big) \nonumber \\
s_{41} &=& {1 \over {d_1 d_2}} \tr_{R_1} \big( T_c T_d T_a T_b \big) 
\tr_{R_2} \big( T_aT_bT_cT_d \big) \nonumber \\
s_{42} &=& -{1 \over {d_1 d_2}}f^{eda} \tr_{R_1} \big( T_bT_cT_d T_e \big) 
\tr_{R_2} \big( T_aT_bT_c \big) \nonumber \\
s_{43} &=& {1 \over {d_1 d_2}} f^{cbd} f^{eag}\tr_{R_1} \big(  T_cT_dT_eT_g
\big)  \tr_{R_2} \big( T_aT_b \big) 
\label{gengroup}
\end{eqnarray}
where we have ommited the superscripts on the generators which label the
representation. In these expressions $d_1$ and $d_2$ are the dimensions of
representations  $R_1$ and $R_2$. 
Their values in the fundamental representations are contained in the following
list:
\begin{eqnarray}
 SU(N)_f: & &   s_{21}= {1\over 4N^2}(N^2-1) \nonumber \\
          & &   s_{31}= {1\over 4N^3}(N^2-1) \nonumber \\
          & &   s_{32}= -{1\over 8N}(N^2-1) \nonumber \\
          & &   s_{41}= {1\over 16N^4}(N^2-1)(N^2+3) \nonumber \\ 
          & &   s_{42}= -{1\over 8N^2}(N^2-1) \nonumber \\
          & &   s_{43}= {1\over 16}(N^2-1) 
\label{casivalsun}
\end{eqnarray}

\begin{eqnarray}
SO(N)_f:  & &   s_{21}= {1\over 8N}(N-1) \nonumber \\
          & &   s_{31}= {1\over 64N}(N-1)(N-2) \nonumber \\
          & &   s_{32}= -{1\over 32N}(N-1)(N-2) \nonumber \\
          & &   s_{41}= {1\over 256N}(N-1)(4-3N +N^2) \nonumber \\ 
          & &   s_{42}= -{1\over 256N}(N-1)(N-2)^2 \nonumber \\
          & &   s_{43}= {1\over 128N}(N-1)(N-2)^2  
\label{casivalson}
\end{eqnarray}

\newpage
     
\vskip 1cm                                               
{\Large{\bf APPENDIX B}}                                 
\vskip .5cm                                              
\renewcommand{\theequation}{\rm{B}.\arabic{equation}}    
\setcounter{equation}{0}                                 

In this appendix we present some elementary facts about semisimple Lie algebras
relevant to the analysis of group factors in the perturbative expansion.

\begin{figure}
\centerline{\hskip.4in \epsffile{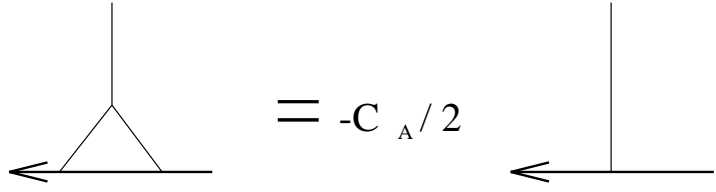}}
\caption{Reduction of a ``fishtail''.}
\label{fishtail.eps}
\end{figure}

Consider the diagram depicted on the left of Fig.~\ref{fishtail.eps}. It is 
possible to reduce its group factor by means of the totally antisymmetry of 
the structure constantss $f_{abc}$ at the cost of introducing $C_A$:

\beq
f_{abc}T_bT_c ={1 \over 2} f_{abc}[T_b,T_c] =
- {1 \over 2} f_{abc}f_{cbd}T_d =-{1 \over 2} C_A T_a. 
\label{identity}
\eeq

There is a point which may be worth commenting. It was already pointed out in
the Appendix B of \cite{numbers} for the  case of knots.
For an arbitrary semisimple Lie algebra
$A=\oplus_{k=1}^n A_k$ in an arbitrary representation, the Wilson line
can be imagined as consisting of $n$ Wilson lines each one corresponding
to one of the simple Lie algebras $A_k$ in its respective
representation. Therefore, a connected diagram  can be regarded as
a sum of similar diagrams where in each term, the legs on the first Wilson
line are attached to some component of it, and
the legs on the second Wilson line to that same component but in the second
Wilson line. The sum runs over all these components. 

When a diagram is
made of some connected subdiagrams, its group factor will be a sum
over all the possible ways of attaching the connected subdiagrams to the $n$
components of the Wilson lines.

As a consequence, the group factors associated to direct products of Lie
algebras cannot be directly identified with the group factors written in
(\ref{gengroup}). For a Lie group $ \otimes_k G_k \otimes U(1)$ with $G_k$ 
simple, the group factors associated with the diagrams in 
figure (\ref{closbasis.eps}) will be:
\begin{eqnarray}
&s_{21}&= \sum_k s_{21}(G_k) + s_{11}^2 \nonumber \\
&s_{31}&= \sum_k s_{31}(G_k) + 3 s_{11}\sum_k s_{21}(G_k) +
 s_{11}^3 \nonumber \\
&s_{32}&= \sum_k s_{32}(G_k) \nonumber \\
&s_{41}&= \sum_k s_{41}(G_k) + 4 s_{11}\sum_k s_{31}(G_k) + 
6 \sum_k \sum_{l<k} s_{21}(G_k)  s_{21}(G_l) + 6s_{11}^2\sum_k s_{21}(G_k)
+  s_{11}^4 \nonumber \\
&s_{42}&= \sum_k s_{42}(G_k) + 2s_{11} \sum_k s_{32}(G_k) \nonumber \\
&s_{43}&= \sum_k s_{43}(G_k) 
\label{semi}
\end{eqnarray}
Of course, if we consider semisimple Lie algebras only, the terms with a 
$s_{11}$ factor will not be present.

For a general diagram, once it has been drawn, its group factor is found as
follows. First one has to obtain its decomposition in a
sum over all the possible ways of attaching the subdiagrams to the $n$
components of the Wilson line, as explained in the previous paragraph.
Then consider a term of the sum, 
reduce all the ``fishtails'' of the diagram chosen by means of
(\ref{identity}). Then, the group
factor of this diagram has to be calculated separately
and written in terms of the basis elements we had chosen by repeated use of
the commutation relations. Repeat this procedure for  each term.
The result is the group factor of the diagram we begun with. Following these steps
for all diagrams present in the sum, we are done.
This procedure is similar, but somehow alternative, to the one presented in
the text, where we begun with open Wilson lines, worked out their group
factor and then closed the lines to loops. The steps carried out to calculate
the group factor for an open general diagram are exactly the ones explained
in \cite{numbers} for the case of knots.

\newpage

\vskip 1cm                                               
{\Large{\bf APPENDIX C}}                                 
\vskip .5cm                                              
\renewcommand{\theequation}{\rm{C}.\arabic{equation}}    
\setcounter{equation}{0}                                 

In this appendix we list the link polynomials which have been used in this work to
compute the numerical invariants introduced in this work up to order four for
links up to six crossings. They were collected from \cite{millet}, after taking
care of our orientation convention.

For $SU(N)$ we have the following list for the HOMFLY polynomials:

\begin{eqnarray}
&2^2_1\,:&\;\;\; z^{-1}( t^{-3} - t^{-1}) - z t^{-1} \nonumber \\
&4^2_{1+}:&\;\;\;  z^{-1}( t^{-5} - t^{-3}) +  z( t^{-5} - 3t^{-3}) - z^3 t^{-3}
\nonumber \\
&4^2_{1-}:&\;\;\;  z^{-1}( t^5 - t^3)  - z( t^3 - t) \nonumber \\
&5^2_{1}\,:&\;\;\;  z^{-1}( t - t^{-1})  + z( t^{-3} - 2t^{-1} + t) -z^3 t^{-1}
\nonumber \\
&6^2_{1+}:&\;\;\; z^{-1}( t^{-5} - t^{-7}) + z( 6t^{-5} - 3t^{-7}) +
z^3( 5t^{-5} - t^{-7}) + z^5 t^{-5} \nonumber \\
&6^2_{1-}:&\;\;\; z^{-1}( -t^{5} - t^{7}) - z( t^5 + t^3 + t) \nonumber \\
&6^2_2\,:&\;\;\; z^{-1}( -t^5 + t^7) + z( t^7 - 2t^5 -2 t^3) - z^3( t^5 - t^3)
\nonumber \\
&6^2_{3+}:&\;\;\; z^{-1}( t^{-3} - t^{-5}) + z( 2t^{-3} - t^{-5} - t^{-7}) 
+ z^3( -t^{-3} + t^{-5}) \nonumber \\
&6^2_{3-}:&\;\;\; z^{-1}( -t^{3} - t^5) + z( -2 t^3 + t^{-1} - t) + z^3 t
\nonumber 
\end{eqnarray}
where $z = q^{1/2}-q^{-1/2}$ and $t = q^{-N/2}$.
\vskip .25cm
And for  $SO(N)$ we have the following list for the Kauffman polynomials:

\begin{eqnarray}
&2^2_1\,:&~ z^{-1} (a^3 - a) + a^2 + z (a^3 - a) \nonumber \\ 
&& \; \nonumber \\ 
&4^2_{1+}:&~  z^{-1}(a^5 - a^3) + a^4 + z (a^6 + 2a^5 - 3a^3)+ z^2 (a^4 - a^6)
+z^3 (a^5 - a^3) 
\nonumber \\                                                      
 && \; \nonumber \\ 
&4^2_{1-}:&~ z^{-1} (a^{-5} - a^{-3}) -a^{-4} +z ( 3a^{-5} - 2a^{-3} -a^{-1})
+ z^2 (-a^{-4} + a^{-2}) + z^3 (a^{-5} - a^{-3}) \nonumber \\
&& \; \nonumber \\ 
&5^2_{1}\,:&~ z^{-1} (a^{-1} - a) - 1 + z ( 2a^{-1} -4a + 2a^{3}) -
z^2 (a^4 + 1) \nonumber \\ 
&& + z^{3} (a^{-1} -3a + 2a^{3}) + z^4 (1 - a^2) \nonumber \\
&& \; \nonumber \\ 
&6^2_{1-}:&~ z^{-1} (-a^{-5} + a^{-7}) -a^{-6} + z (6a^{-7} -4a^{-5} - a^{-3}
-a^{-1})
\nonumber \\ 
&&+z^2 (-3a^{-6} +2a^{-4} + a^{-2}) + z^3 (5a^{-7} -4a^{-5} - a^{-3} ) -
z^4 (a^{-6} + a^{-4}) + z^5 ( -a^{-5} + a^{-7} ) \nonumber \\
&& \; \nonumber \\ 
&6^2_{1+}:&~ z^{-1} (a^5 - a^7) -a^6 + z (6a^5 - 4a^7 -a^9 -a^{11}) +
z^2 (-3a^6 + 2a^8 -a^9 + a^{10}) \nonumber \\
&& + z^3 (5a^5 - 4a^7 -a^9 ) 
+ z ( - a^6 + a^8 ) + z^5 (a^5 -a^7) \nonumber \\
&& \; \nonumber \\ 
&6^2_2\,:&~ z^{-1} ( -a^{-5} - a^{-7} ) -a^{-6} + z(2a^{-9}+ 3a^{-7} - 3a^{-5}
-2a^{-3}) 
 + z^2 (-2a^{-6} + a^{-8} + a^{-4} ) \nonumber \\ && +
z^3 (a^{-9} -a^{-7} - 2a^{-5} -a^{-3}) + z^4 (a^{-4} -2a^{-6} - a^{-8})
+ z^5 ( -a^{-5} + a^{-7} ) \nonumber \\
&& \; \nonumber \\ 
&6^2_{3+}:&~ z^{-1} (a^3 - a^5) -a^4 + z (2a^3 - a^5 - a^9) +z^2 (-3a^6 + 3a^8)
\nonumber \\
&& + z^3 (a^3 - a^9) + z^4 (a^4 - 3a^6 +2a^8) + z^5 (a^5 - a^7) \nonumber \\
&& \; \nonumber \\ 
&6^2_{3-}:&~ z^{-1} (a^{-5} - a^{-3}) -a^{-4} + z (2a^{-5} - a^{-3} -a) +
z^2 (-3a^{-2} + 3) \nonumber \\
&& + z^3 (a^{-5} - a) + z^4 (a^{-4} - a^{-2} + 2) +
z^5 (a^{-1} - a^{-3}) \nonumber
\end{eqnarray}
where $z = q^{1/4} - q^{-1/4}$ and $a = q^{N-1 \over 4}$.


\end{document}